\documentclass[pra,english,preprintnumbers,amsmath,amssymb,nofootinbib,twocolumn,superscriptaddress]{revtex4-1}

\pdfoutput=1

\usepackage[latin1]{inputenc}
\usepackage{graphicx}
\usepackage{color}
\usepackage{bbm}
\usepackage{amssymb}
\usepackage{amsmath}
\usepackage{tabularx}

\usepackage{dsfont}

\def\0#1#2{\frac{#1}{#2}}

\def\s0#1#2{\mbox{\small{$ \frac{#1}{#2} $}}}

\newcommand{\beq}{\begin{equation}}
\newcommand{\eeq}{\end{equation}}
\newcommand{\bea}{\begin{eqnarray}}
\newcommand{\eea}{\end{eqnarray}}
\newcommand{\tr}{\mathrm{Tr}}

\newcommand{\hi}{h_{\rm I}}
\newcommand{\afit}{{\mathcal A}}
\newcommand{\bfit}{{\mathcal B}}
\newcommand{\cfit}{{\mathcal C}}
\newcommand{\be}{\begin{eqnarray}}
\newcommand{\ee}{\end{eqnarray}}

\usepackage{babel}
\makeatother
\begin{document}

\title{Thermal equation of state of polarized fermions in one dimension \\ via complex chemical potentials}

\author{Andrew C. Loheac} 
\affiliation{Department of Physics and Astronomy, University of North Carolina, Chapel Hill, North Carolina 27599, USA}

\author{Jens Braun} 
\affiliation{Institut f\"ur Kernphysik (Theoriezentrum), Technische Universit\"at Darmstadt, 
D-64289 Darmstadt, Germany}

\author{Joaqu\'{i}n E. Drut} 
\affiliation{Department of Physics and Astronomy, University of North Carolina, Chapel Hill, North Carolina 27599, USA}

\author{Dietrich Roscher}
\affiliation{Institut f\"ur Kernphysik (Theoriezentrum), Technische Universit\"at Darmstadt, 
D-64289 Darmstadt, Germany}

\begin{abstract}
We present a nonperturbative computation of the equation of state of polarized, attractively interacting, 
nonrelativistic fermions in one spatial dimension at finite temperature.
We show results for the density, spin magnetization, magnetic susceptibility, and Tan's contact. We compare with
the second-order virial expansion, a next-to-leading-order lattice perturbation theory calculation,
and interpret our results in terms of pairing correlations. Our lattice Monte Carlo calculations 
implement an imaginary chemical potential difference to avoid the sign problem. The thermodynamic results on 
the imaginary side are analytically continued to obtain results on the real axis. We focus on an 
intermediate- to strong-coupling regime, and cover a wide range of temperatures and spin imbalances.
\end{abstract}

\maketitle
\section{Introduction} 
In the last two decades, experimental studies with ultracold atomic gases have made consistent strides 
towards the increasingly clean and controlled characterization of strongly coupled matter,
particularly in nonperturbative regimes that are out of reach for conventional theory methods~\cite{RevExp}.
This represents a challenge to the theory side, which has been met in some cases but remains open in general.

As is well known, spatial dimensionality plays a crucial role in these systems regardless of the strength of the interaction.
Although it is also generally understood that interactions tend to dominate in lower dimensions and, conversely,
mean-field descriptions become more reliable in higher dimensions, {they in general do not allow for quantitative predictions. 
This can be viewed as a signal that fluctuation effects still play a prominent role, calling for more sophisticated theoretical tools.}
Both from the theory and experiment sides, considerable progress has been made in the study of three-dimensional (3D) systems in a variety of situations 
(e.g., in harmonic traps, homogeneous space, polarized, unpolarized, in the ground state, at finite temperature, etc.; see 
Ref.~\cite{RevTheory} for reviews), 
in particular with emphasis on the so-called BEC-BCS crossover and scale-invariant regimes~\cite{ZwergerBook}, as well as the Efimov 
effect~\cite{EfimovBraaten}. In recent years, there has also been increasing activity in similar studies in 2D (see, e.g., Ref.~\cite{BECBCS2D} 
for a recent review), where the possibility of accessing directly the superfluid Berezinskii-Kosterlitz-Thouless transition~\cite{BKT}
has been a major drive.

In this context, the motivation to study one-dimensional systems, in particular fermions, is manifold. 
Large classes of 1D problems can be solved exactly {at zero temperature} via powerful techniques such as the Bethe ansatz, which has
propelled a fair amount of work over the last few decades (see, e.g., Refs.~\cite{OneDReview,OneDBooks}).
However, a new wave of interest has been underway for a few years. This renewed activity stems in part 
from the realization of 1D systems in the form of atomic gases in highly constrained, quasi-1D optical traps, 
but it is also due to the advent of quantum-information concepts in condensed-matter physics~\cite{EEReview} and their
connection to quantum phase transitions (in particular topological ones) in low dimensions.

Systems of spin-polarized, attractively interacting fermions are particularly appealing because of the 
potential occurrence of spontaneously broken translation invariance in exotic superfluid {phases~\cite{LOFF}.} 
The focus of the present work, however, is on the basic thermodynamic equations of state of spin-polarized fermions in 1D, rather than on
detecting potential exotic superfluid phases.
We compute the density and the spin magnetization as well as Tan's contact, which encodes the importance
of high-momentum correlations in systems with short-range interactions~\cite{TanContact,ContactReview}.
Our study is distinguished from previous ones in that we make use of complex chemical potentials to overcome 
the so-called sign problem. The latter has been a major roadblock in lattice Monte Carlo calculations of asymmetric 
systems (e.g., systems with mass or spin imbalance~\cite{BraunEtAl,ImaginaryMassesMC}, or at finite quark density in the 
case of QCD~\cite{QCDImbalance}), 
as explained below. As a proof of principle, we consider spin-$1/2$ fermions in 1D with attractive contact interactions, but
more general situations including richer interactions and higher dimensions can be studied with the same methods.
Note that we do not determine the exact nature of the ground state of the theory in our present study;
as we explain in detail below, our approach based on complex chemical potentials is in fact unable to reach the ground state
(and at fixed temperature is not able to reach arbitrary polarizations).
Nevertheless, the ground state indirectly leaves its imprint in our computation of the basic thermodynamic equations of state.

The Hamiltonian we analyze is that of the Gaudin-Yang model~\cite{GaudinYang},
\beq
\hat H = -\frac{\hbar^2}{2m}\sum_i \nabla_{i}^2 - \sum_{i < j} g\delta(x^{}_i-x^{}_j),
\eeq
where the sums are over all particles.
In our study, we will consider polarized systems, in the sense that they will have a 
nonvanishing average spin magnetization in general. In the grand-canonical ensemble, the partition function of
such a system is
\beq
\label{Eq:PartitionZ}
\mathcal Z(\beta\mu^{}_\uparrow,\beta\mu^{}_\downarrow) = \text{Tr} \exp \left[{-\beta( \hat H - \mu^{}_\uparrow \hat N^{}_\uparrow - \mu^{}_\downarrow \hat N^{}_\downarrow)}\right],
\eeq
where $\mu^{}_s$ is the chemical potential for spin $s=\uparrow,\downarrow$ particles, $\hat N^{}_s$ is the
corresponding particle number operator, and $\beta$ is the inverse temperature.

As mentioned above, asymmetric systems
are challenging for lattice Monte Carlo calculations due to the appearance of the sign problem.
To circumvent this difficulty, we implement the approach put forward in Ref.~\cite{BraunEtAl}. 
Here we attempt this type of calculation for a nonrelativistic 
theory; a similar strategy has been applied by some of the present authors to 
ground-state calculations in the mass-imbalanced case~\cite{ImaginaryMassesMC}.
In the present case, we take the chemical potential for each fermion species to be complex, 
but such that one is the complex conjugate of the other: $\mu^{}_\uparrow = \mu^{*}_\downarrow$. As a consequence,
the fermion determinants in the Monte Carlo calculation (see, e.g., Ref.~\cite{ImaginaryMassesMC, MCReviews}) are also 
complex conjugates of one another, and the probability measure is thus non-negative. 

In this approach, the overall chemical potential
${\mu} = (\mu^{}_\uparrow + \mu^{}_\downarrow)/2$ is real, as usual, but the asymmetry parameter 
{$h = (\mu^{}_\uparrow - \mu^{}_\downarrow)/2$ is imaginary. For convenience, we define $\hi:=\text{Im}\,h=-ih$, with $h_{\rm I}$ being a real-valued number. 
The total density 
$n = n^{}_\uparrow+n^{}_\downarrow$ in our study based on the grand-canonical ensemble 
is then still a real-valued number, while the so-called spin magnetization $m = n^{}_\uparrow - n^{}_\downarrow$
is imaginary.} {These, as well as every output of the calculation, must be analytically continued to the real-$h$ axis
in order to obtain the physical results. While the analytic continuation procedure introduces some degree of arbitrariness
in the final results (see below), it should be pointed out that the results on the imaginary side are fully nonperturbative and,
in principle, exact, and certain aspects of the functional dependence with respect to $h$ are known.
Note, for instance, that the asymmetry $h$ always enters in the 
calculations as a function of $\beta h$.} Moreover, it can be shown that the results for imaginary asymmetries will be {2$\pi$ periodic} in $\beta \hi$;
see Ref.~\cite{BraunEtAl}.
This is therefore a compact parameter and we will restrict it to the interval $[-\pi,\pi]$. The symmetry
in the form of the partition function under spin exchange indicates that the physics it describes is independent
of the sign of $h$, i.e., we expect our results to be either odd or even functions of $h$, depending on the observable. 

We would like to 
emphasize that although it is, in principle, exact, our present Monte Carlo approach to spin-imbalanced Fermi gases is not capable of studying the zero-temperature
limit, but is limited to finite temperatures $|\beta\hi| \leq \pi$ {being a direct consequence of the $2\pi$ periodicity in $\beta\hi$.
For the same reason, one may say that at fixed temperature (i.e. $\beta$) not all polarizations are achievable, as the above constraint 
limits the range of $\hi$.
In this sense, our present approach is complementary to the {\it Bethe} ansatz~\cite{OneDReview,OneDBooks} which 
allows for an exact solution of one-dimensional Fermi gases in the zero-temperature limit. Still, our approach enables us to compute the 
finite-temperature equation of state in a certain parameter range which can then potentially be compared to experiments, 
see, e.g., Ref.~\cite{Experiments1D}.} {The present, however, should rather be regarded as a proof-of-principle
work that serves as an intermediate step towards calculations in higher dimensions.}

It should also be pointed out that we do not expect phase transitions to be present in this 1D system as a function of temperature
{due to the absence of spontaneous symmetry {breaking~\cite{MWHC}.}
This fact should allow for a more reliable analytic continuation from {$\beta \hi$ to} the 
real side. Conversely, the appearance of accumulation points of zeros of the partition function (the so-called 
Yang-Lee zeros) in higher dimensions may complicate the analytic continuation in those cases.

\section{Scales and computational technique}
\subsection{General setup}

The problem of fermions with a contact interaction is ultraviolet-finite in one spatial dimension. Therefore,
the bare coupling has a physical meaning: in the continuum limit, $g = 2/a_0$, where $a_0$ is the scattering length
for the symmetric channel (see e.g.~\cite{ScatteringIn1D}); accordingly, $g$ will be reported in units of $\sqrt{\beta}$.
Note that the thermal de Broglie wavelength is $\lambda^{}_T = \sqrt{2 \pi \beta}$.

To characterize the thermodynamics of polarized interacting fermions in one dimension we compute three quantities,
namely the {density $n$, the spin magnetization $m$ and} the contact $C$, as functions of the 
inverse temperature $\beta$, the average chemical potential $\mu$, the chemical potential difference $h$, and the 
coupling $g$. To make all of these quantities dimensionless, we utilize the noninteracting unpolarized density $n^{}_0$
as the scale for $n$ and $m$, i.e. we report $n/n_0^{}$ and $m/n^{}_0$.
On the other hand, for the input parameters we set $\beta$ as the main scale, i.e. we report the physical
results as functions of 
\beq
\beta {\mu}, \ \ \ \beta h, \ \ \ \text{and}\ \ \ \ \lambda \equiv \sqrt{\beta} g\,,
\eeq
where the contact $C$ is made dimensionless by $C_0$, the unpolarized result at $\beta{\mu} = 0$. {Note that, since we 
use an imaginary chemical potential difference in our Monte Carlo studies, the ``bare" outcome of these simulations will be given as a function
of $\beta{\mu}$, $\beta\hi$, and $\lambda$.}

From the density and the magnetization one may obtain the isothermal compressibility and magnetic susceptibility simply
by taking derivatives with respect to $\beta \mu$ and $\beta h$, respectively. Mixed response functions (of $n$ with respect to
$\beta h$, or $m$ with respect to $\beta \mu$) may be obtained in the same fashion. On the other hand, numerical
integration of $n$ with respect to $\beta \mu$ provides the pressure for each value of $\beta h$.

The computational method utilized in this work is very similar to the one of Refs.~\cite{BDM1,BDM2,EoSUFG2}, but reduced to one spatial
dimension and generalized to complex chemical potentials as explained above (see also Ref.~\cite{EOS1D}).
Because one-dimensional problems are computationally inexpensive, it is possible to calculate in very large lattices, e.g. 
{$N_x^{} \sim {\mathcal O}(10^2)$.} For such sizes, the continuum limit is easily achieved by lowering the density, while
still remaining in the many-particle (i.e. thermodynamic) regime. For the proof-of-principle calculations presented here,
we fix $\lambda=1.0$. This was chosen as being in the intermediate-to-strong-coupling regime,
which is typically outside the range of {validity of perturbative approaches.} For such a coupling strength, a lattice size 
of $N_x^{}=61$, which we fixed throughout this study, is sufficient to provide a good quantitative understanding of 
the continuum limit (see Ref.~\cite{EOS1D}). The physical extent of the system is $L = N_x^{} \ell$, where $\ell=1$ fixes the
spatial lattice units. The extent of the temporal lattice is given by $\beta = \tau N_\tau$, where we
take $\tau = 0.05/\ell^2$.

\subsection{Computing the density and magnetization at imaginary asymmetry}

At imaginary chemical potential asymmetry, the partition function of Eq.~(\ref{Eq:PartitionZ}) can be written 
in terms of a Hubbard-Stratonovich auxiliary field $\sigma$ as
\beq
\mathcal Z = \int \mathcal D \sigma |\det (1 + z U[\sigma])|^2\,,
\eeq
where $z \equiv z^{}_\uparrow = \exp(\beta \mu^{}_\uparrow) = \exp(\beta \mu^{*}_\downarrow) = z^{*}_\downarrow$,
and $U[\sigma]$ is a matrix that encodes the dynamics of the system (see e.g. Ref.~\cite{MCReviews}).
We thus identify 
\beq
P[\sigma] \equiv |\det (1 + z U[\sigma])|^2
\eeq
as the non-negative probability measure for our Monte Carlo calculations. The total (average) particle {density $n$ is} then obtained 
in those calculations {using
\beq
n = \frac{1}{L}\frac{\partial \ln \mathcal Z}{\partial(\beta \mu)} = 
\frac{2}{\mathcal ZL}  \int \mathcal D \sigma P[\sigma] \text{Re}\left[ \tr \left( \frac{z U[\sigma]}{1 + zU[\sigma]}    \right) \right],\label{Eq:NDefinition}
\eeq
and} the (average) spin magnetization $m$ can similarly be {calculated using $m=(1/L)\partial \ln {\mathcal Z}/\partial (\beta h)$. To circumvent the
sign problem in our Monte Carlo simulations, however, we rather compute
\beq
m = \frac{1}{L}\frac{\partial \ln \mathcal Z}{\partial(\beta \hi)} = 
-\frac{2i}{\mathcal ZL}  \int \mathcal D \sigma P[\sigma] \text{Im}\left[ \tr \left( \frac{z U[\sigma]}{1 + zU[\sigma]}    \right) \right]. \label{Eq:MDefinition}
\eeq
Since we assume that $m$ is analytic as a function of general complex-valued $h$, at least in a finite domain about $h=0$, we shall use the same
label as for the physical spin magnetization.}

To determine the contact, we use the same approach as in Ref.~\cite{EOS1D}.
The definition in 1D is
\beq
{C}\equiv \frac{2}{\beta \lambda^{}_T}\left . \frac{\partial (\beta \Omega)}{\partial (a^{}_{0}/\lambda^{}_T)} \right |^{}_{\mu,T},
\eeq
where $\Omega = -\frac{1}{\beta}\ln \mathcal Z$ is the grand thermodynamic potential. Using the Feynman-Hellman theorem, 
it can be shown that
\beq
{C} = -{g} {\langle \hat V \rangle},
\eeq
where ${\langle \hat V \rangle}$ is the thermal expectation value of the {interaction operator}. The latter
can be {computed} in Monte Carlo calculations using derivatives of $\ln \mathcal Z$ with respect to the 
bare lattice coupling $g$ or the lattice spacing $\tau$.

\section{Results}
\subsection{Monte Carlo results for imaginary asymmetry}
Throughout this work, we present Monte Carlo calculations at $\beta = 8.0$ and a lattice size of $N^{}_x = 61$.
For the purposes of demonstrating the method, we fix the dimensionless coupling to $\lambda = 1.0$, although a variety 
of coupling strengths may also be explored using the same technique.
\begin{figure}[t]
  \centering
    \includegraphics[width=1.03\columnwidth]{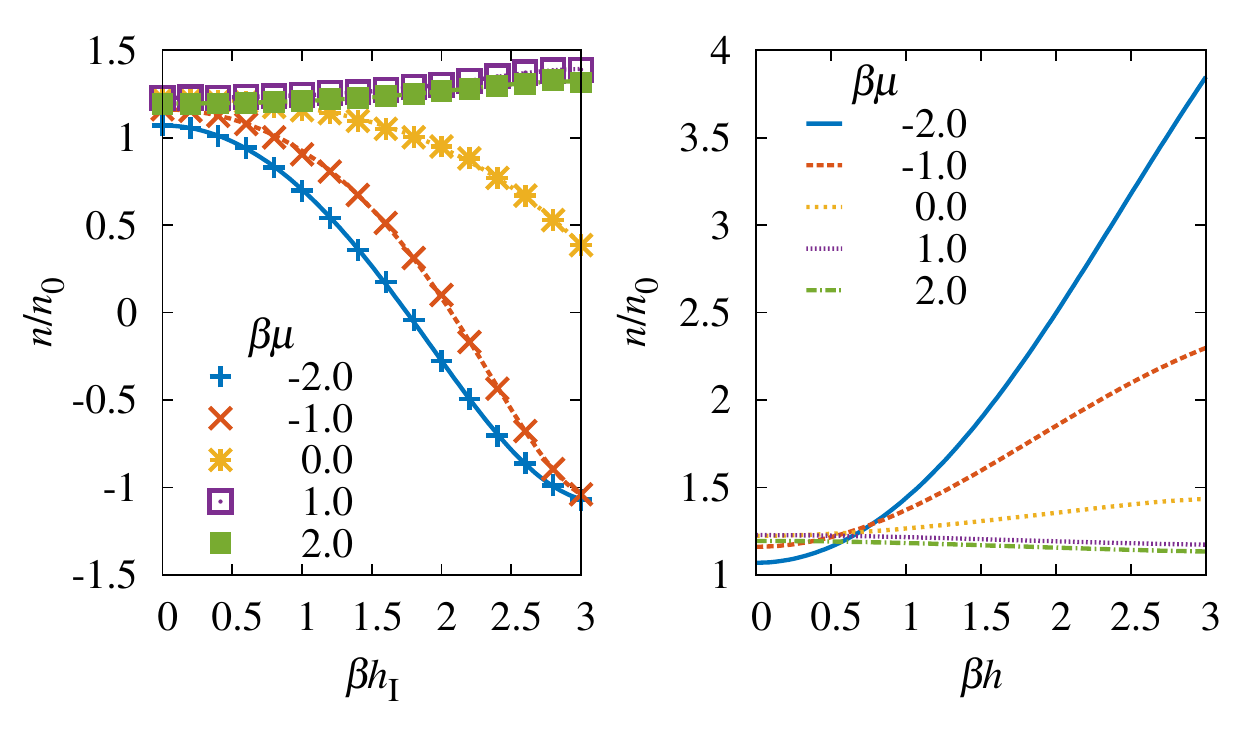}
     \caption{\label{Fig:DensityIM} (Color online) Left: Density as a function of the imaginary chemical potential {difference $\beta \hi$}
     at various values of $\beta\mu$ for a dimensionless coupling of $\lambda = 1.0$. Right: Analytic continuation of the density as a function of $\beta h$ at various values of $\beta\mu$. In both plots the density is an even function about the origin and is plotted in units of the density of the noninteracting, unpolarized system.}
\end{figure}
\begin{figure}[t]
  \centering
    \includegraphics[width=1.03\columnwidth]{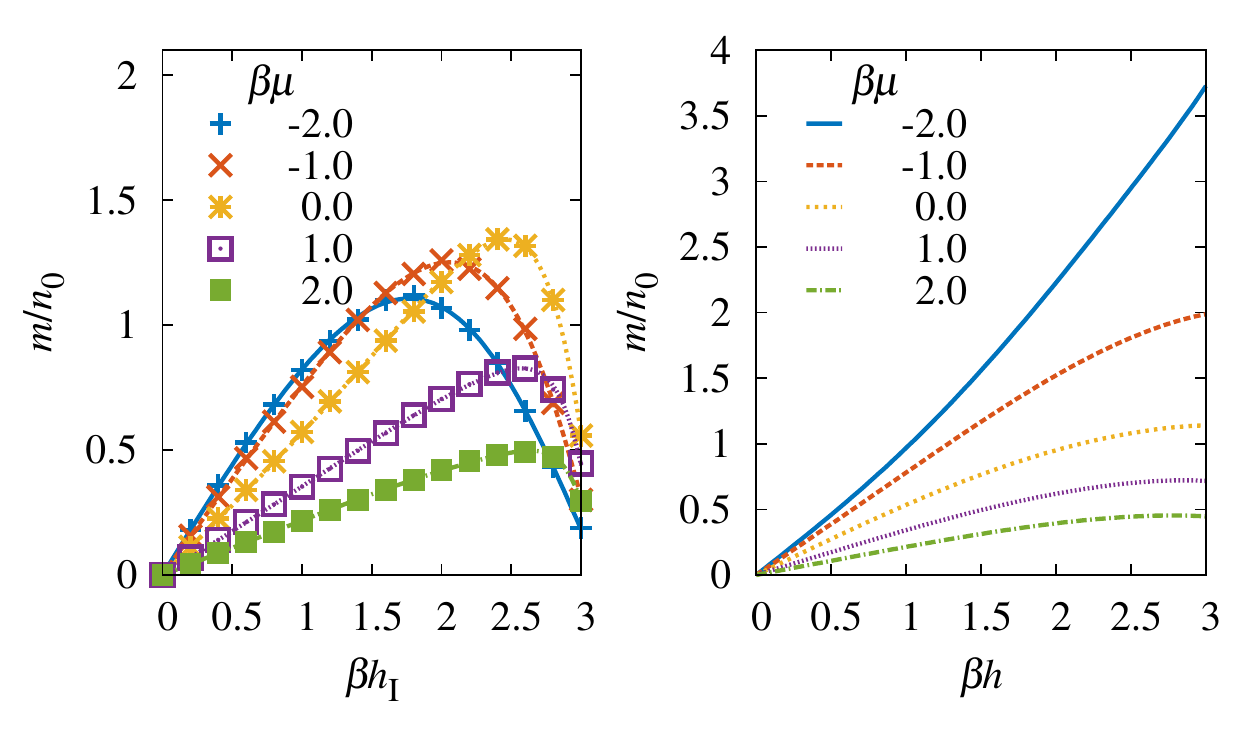}
     \caption{\label{Fig:PolarizationIM} (Color online) Left: Magnetization as a function of the imaginary chemical potential difference {$\beta \hi$} 
     at various values of $\beta\mu$ for a dimensionless coupling of $\lambda = 1.0$. Right: Analytic continuation of the magnetization as a function 
     of $\beta h$ at various values of $\beta\mu$. In both plots the magnetization is an odd function about the origin and is plotted in units of the
     density of the noninteracting, unpolarized system.}
\end{figure}
The imaginary asymmetry {parameter $\beta \hi$} was varied over a full period $[-\pi,\pi]$, and the chemical potential parameter 
{$\beta {\mu}$ was} varied in the interval $[-4.0, 4.0]$, covering the semiclassical regime (where the virial expansion is valid) 
to the fully quantum mechanical regime. For each point in the plots below, we have taken 1000 de-correlated Monte Carlo 
samples, thus ensuring that the statistical uncertainty is below $10\%$.

In Fig.~\ref{Fig:DensityIM} we show the density as a function {of $\beta \hi$ and $\beta h$}, respectively, for representative values of $\beta\mu$.
Similarly, Fig.~\ref{Fig:PolarizationIM} displays the magnetization and Fig.~\ref{Fig:ContactIM} shows Tan's contact. The statistical
error of the Monte Carlo calculations on the imaginary side is estimated to be on the order of the symbol size in all three figures.
In all cases we show the Monte Carlo results at {$\beta \hi$} on the left panel, and the corresponding
analytic continuation (described in detail below) on the right.
Although the imaginary side of the problem is not physically meaningful, the results are non-perturbative, and it
is reassuring that the data falls on smooth curves that respect the even or odd symmetry {around $\beta \hi=0$.
We therefore display only the positive interval $\beta \hi \in [0,\pi]$.}

\begin{figure}[t]
  \centering
    \includegraphics[width=1.03\columnwidth]{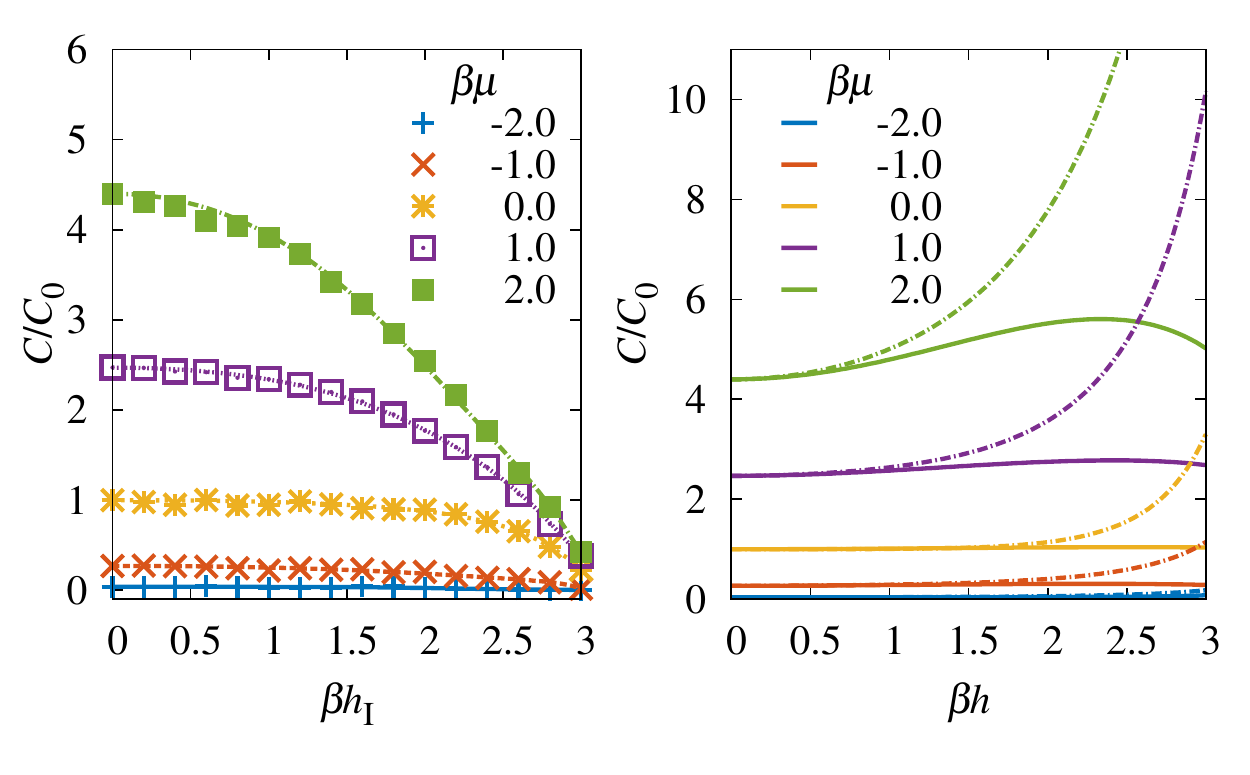}
     \caption{\label{Fig:ContactIM} (Color online) Left: Tan's contact as a function of the imaginary chemical potential difference {$\beta \hi$ at} 
     various values of $\beta\mu$ for a dimensionless coupling of $\lambda = 1.0$. 
     $C^{}_0$ is the contact at 
 $\beta h = \beta\mu = 0$. Right: Analytic continuation of the contact as a function of $\beta h$ at various 
 values of $\beta \mu$. In both plots the contact is an even function about the origin. The curves on the right are {color-wise paired by their value of $\beta\mu$, which
 coincides with the value on the left (by color code, or from top to bottom). Dashed-dotted lines give the results from a fit of the data to 
 {the} polynomial-type ansatz~\eqref{eq:poltype}, whereas solid lines
 result from fits of the data to our Pad\'{e}-type ansatz~\eqref{Eq:DensityFit} for the fit functions.}}
\end{figure}

\subsection{Analytic continuation to the real axis} \label{Sec:AnalyticContinuation}
{In order to obtain the results of physical interest, we need to analytically continue the data
from our Monte Carlo study to real-valued chemical potential differences. 
To this end, we fit our data to a specific ansatz for, e.g., the spin magnetization.
Clearly, this is a critical step as the functional form of the ansatz is {\it a priori} unknown. However, as already discussed above,
we know that, depending on the observable, the ansatz must be either an odd or even function in $\beta\hi$ and that the
partition function is periodic in $\beta\hi$. Moreover, the virial expansion of the partition function $\mathcal Z$
suggests that $\mathcal Z$ can be written as a (asymptotic) series in powers of $\cos(\beta\hi)$, see
also our discussion in Sect.~\ref{sec:other}. From such an analysis of the virial expansion, it also follows that higher-order 
terms in $\cos(\beta\hi)$ become particularly important for $\beta\mu \gg 1$. In order to take such higher order terms
effectively into account, an ansatz of the type
\be
\sim \frac{1}{1+\cos(\beta\hi)}
\ee
may be considered appropriate as it can be rewritten as an asymptotic series in powers of $\cos(\beta\hi)$, which still obeys the periodicity
in $\beta\hi$. An ansatz of this type may be viewed as a {\it Pad\'{e}} approximant of cosines.}

{To be specific, we choose to fit our Monte Carlo data for the magnetization
at finite $\beta\hi$ with the {function
\beq
{f(\beta \hi) = -i\frac{\afit\sin(\beta \hi)}{1+\bfit\cos(\beta \hi) + \cfit F(\beta \hi)}}
\label{Eq:PolarizationFit}
\eeq
where} $\afit$, $\bfit$, and $\cfit$ are free real-valued fit parameters and the function $F$ is given by 
\beq
F(x) = \cos\left(x^3/\pi^2\right).
\eeq
Note that this last function has been found empirically. As written, 
it is analytic on the whole complex plane, in particular on a disk of radius $\pi$ centered at the origin. However, it is
not periodic in $\beta \hi$, as we required above. Here, we have indeed given up this constraint as we found that
a large number of fit parameters is required to meet this criterion, rendering the fitting algorithm potentially unstable. For example, an ansatz
of the form $\sim {\sum_{k=1}\afit_k\sin(k\beta \hi)}/(1+{\sum_{k=1}\bfit_{k}\cos(k\beta \hi)})$ is compatible with the above-mentioned constraints. However, the fit is of much lower quality than those obtained using Eq.~\eqref{Eq:PolarizationFit}.
Interestingly, the parameter~$\cfit$ associated with the function $F$ is in most cases found to 
be small when we fit our Monte Carlo data with the ansatz~\eqref{Eq:PolarizationFit}, see Tab.~\ref{Table:PolarizationFitParameters}.}

The analytic continuation $\tilde{f}(\beta h)$ of Eq.~(\ref{Eq:PolarizationFit}) is obtained {simply by setting $\hi= -i h$, and is therefore given {by
\beq
\tilde{f}(\beta h) = -\frac{\afit\sinh(\beta h)}{1+\bfit\cosh(\beta h) + \cfit{\tilde F}(\beta h)}\,,
\label{Eq:PolarizationFitAC}
\eeq
where} ${\tilde F}$ is the analytic continuation of $F$.}

In the case of the density and the contact, on the other hand, we expect even functions of $\beta \hi$, and therefore 
we have chosen to fit the {function
\beq
g(\beta \hi) = \gamma\left[\frac{1+\afit\eta(\beta \hi)}{1+\bfit\eta(\beta \hi) + \cfit\eta((\beta \hi)^3/\pi^2)}\right],
\label{Eq:DensityFit}
\eeq
where} $\eta(x) = 1-\cos(x)$. Here, $\gamma$ is {a} parameter that is fixed by the exactly known value at {$\beta h = \beta\hi=0$, while $\afit$, 
$\bfit$ and $\cfit$ are free real-valued fit parameters.} Once the parameters are
obtained, the analytic continuation $\tilde{g}(\beta h)$ of Eq.~(\ref{Eq:DensityFit}) is given {by
\beq
\tilde{g}(\beta h) = \gamma\left[\frac{1+\afit\tilde{\eta}(\beta h)}{1+\bfit\tilde{\eta}(\beta h)+\cfit\tilde{\eta}((\beta h)^3/\pi^2)}\right], \label{Eq:DensityFitAC}
\eeq
where} $\tilde{\eta}(x) = 1-\cosh(x)$.

The fit parameters for the density, magnetization, and Tan's contact are provided in Tables~\ref{Table:DensityFitParameters}}, 
\ref{Table:PolarizationFitParameters}, and \ref{Table:ContactFitParameters}, respectively. Since the fits to the Monte Carlo data are sensitive to initial 
parameter values, the fitting algorithm performs several such fits with random initial parameter values in the interval $[-1,1]$, and the 
best fit with the minimum mean residuals is chosen for the analytic continuation. Given the functional form of Eqs.~(\ref{Eq:PolarizationFit}) and~(\ref{Eq:DensityFit}), one should consider that poles may appear in 
either the fit or analytic continuation for a given set of fit parameters. {Since we expect these quantities to be analytic for one dimensional Fermi gases},\footnote{{In higher dimensions, 
phase transitions may occur, potentially rendering physical quantities non-analytic
at the transition point. Depending on the observable and the employed ansatz, such poles may therefore not just be an artifact resulting from the details of the fitting procedure but may be a hint to an underlying physical effect.}} 
the fitting algorithm eliminates any fits that demonstrate such behavior. Of course other functional forms may be considered for the analytic
 continuation with consideration to the constraints discussed. 

\begin{table}[t]
\begin{center}
\caption{\label{Table:DensityFitParameters}
Fit parameters for the density as they appear in Eq. \eqref{Eq:DensityFit} at a constant 
dimensionless coupling of $\lambda = 1.0$ for various values of $\beta\mu$, as well as the $\chi^2$ per degree of freedom for each fit. 
Note that $\gamma$ is not a fit parameter (see main text).
The value in parentheses indicates the calculated uncertainty of the least significant digit for each fit parameter.}
\begin{tabularx}{\columnwidth}{@{\extracolsep{\fill}}c c c c c c}
\hline\hline
$\beta\mu$ & $\gamma$ & ${\afit}$ & ${\bfit}$ & ${\cfit}$ & $\chi^2$\\
\hline
-3.6& 1.0080 & -0.9670(4) & -0.041(1) & -0.001(1) & 0.71\\
-3.2& 1.0172 & -0.9511(4) & -0.0565(9) & -0.002(1) & 0.58\\
-2.8& 1.0294 & -0.9281(4) & -0.080(1) & -0.006(1) & 0.54\\
-2.4& 1.0461 & -0.8942(9) & -0.118(3) & 0.005(3) & 1.29\\
-2.0& 1.0698 & -0.848(1) & -0.156(3) & -0.000(4) & 1.66\\
-1.6& 1.1026 & -0.7904(6) & -0.198(2) & -0.016(2) & 0.35\\
-1.2& 1.1245 & -0.716(1) & -0.276(3) & -0.008(4) & 1.25\\
-0.8& 1.1731 & -0.6334(6) & -0.323(2) & -0.019(3) & 0.63\\
-0.4& 1.1921 & -0.5433(9) & -0.364(4) & -0.034(7) & 1.30\\
0.0& 1.2240 & -0.492(2) & -0.415(2) & -0.056(5) & 2.74\\
0.4& 1.2359 & -0.503(3) & -0.493(3) & -0.009(1) & 5.21\\
0.8& 1.2361 & -0.4(1) & -0.4(1) & 0.00(1) & 0.54\\
1.2& 1.2214 & -0.18(5) & -0.22(5) & -0.013(6) & 0.97\\
1.6& 1.2109 & -0.32(2) & -0.34(2) & 0.004(1) & 0.71\\
2.0& 1.1948 & -0.33(3) & -0.35(3) & 0.005(2) & 1.81\\
2.4& 1.1763 & -0.22(2) & -0.24(2) & -0.002(1) & 0.41\\
2.8& 1.1625 & -0.11(3) & -0.13(3) & -0.007(1) & 0.43\\
3.2& 1.1507 & -0.20(3) & -0.22(3) & -0.001(1) & 0.45\\
3.6& 1.1397 & -0.07(6) & -0.08(6) & -0.009(3) & 0.60\\
\hline\hline
\end{tabularx}
\end{center}
\end{table}

\begin{table}[t]
\begin{center}
\caption{\label{Table:PolarizationFitParameters}
Fit parameters for the magnetization as they appear in Eq. \eqref{Eq:PolarizationFit} at a constant dimensionless 
coupling of $\lambda = 1.0$ for various values of $\beta\mu$, as well as the $\chi^2$ per degree of freedom for each fit.
The value in parentheses indicates the calculated uncertainty of the least significant digit for each fit parameter.}
\begin{tabularx}{\columnwidth}{@{\extracolsep{\fill}}c c c c c}
\hline\hline
$\beta\mu$ & ${\afit}$ & ${\bfit}$ & ${\cfit}$ & $\chi^2$\\
\hline
-3.6 & 1.022(4) & 0.044(3) & 0.011(5) & 0.48\\
-3.2 & 1.031(4) & 0.062(3) & 0.011(5) & 0.50\\
-2.8 & 1.048(3) & 0.094(3) & 0.012(4) & 0.41\\
-2.4 & 1.047(6) & 0.146(5) & -0.015(6) & 1.01\\
-2.0 & 1.062(5) & 0.202(4) & -0.023(6) & 1.35\\
-1.6 & 1.098(4) & 0.269(3) & 0.000(4) & 0.48\\
-1.2 & 1.134(4) & 0.387(3) & 0.017(4) & 1.80\\
-0.8 & 1.142(3) & 0.521(3) & 0.037(3) & 5.81\\
-0.4 & 1.089(4) & 0.632(3) & 0.052(3) & 8.90\\
0.0 & 0.978(3) & 0.702(2) & 0.067(2) & 23.32\\
0.4 & 0.833(3) & 0.711(3) & 0.088(3) & 36.24\\
0.8 & 0.695(3) & 0.684(3) & 0.123(3) & 7.36\\
1.2 & 0.574(2) & 0.660(3) & 0.156(3) & 4.95\\
1.6 & 0.473(2) & 0.651(3) & 0.175(3) & 7.57\\
2.0 & 0.397(2) & 0.639(4) & 0.193(4) & 8.02\\
2.4 & 0.340(2) & 0.630(5) & 0.214(5) & 11.10\\
2.8 & 0.292(2) & 0.634(6) & 0.215(6) & 14.55\\
3.2 & 0.256(1) & 0.638(5) & 0.218(4) & 14.73\\
3.6 & 0.227(2) & 0.636(7) & 0.218(7) & 24.89\\
\hline\hline
\end{tabularx}
\end{center}
\end{table}

\begin{table}[t]
\begin{center}
\caption{\label{Table:ContactFitParameters}
Fit parameters for the contact as they appear in Eq. \eqref{Eq:DensityFit} at a constant dimensionless coupling of $\lambda = 1.0$ for 
various values of $\beta\mu$, as well as the $\chi^2$ per degree of freedom for each fit. Note that $\gamma$ is not a fit parameter (see main text).
The value in parentheses indicates the calculated uncertainty of the least significant digit for each fit parameter.}
\begin{tabularx}{\columnwidth}{@{\extracolsep{\fill}}c c c c c c}
\hline\hline
$\beta\mu$ & $\gamma$ & ${\afit}$ & ${\bfit}$ & ${\cfit}$ & $\chi^2$\\
\hline
-3.6& 0.0003 & -0.4(2) & -0.60(4) & 0.15(4) & 0.47\\
-3.2& 0.0019 & -0.50(5) & -0.50(3) & -0.00(5) & 0.52\\
-2.8& 0.0062 & -0.503(2) & -0.10(5) & -0.40(8) & 0.12\\
-2.4& 0.0159 & -0.5(1) & -0.40(9) & 0.4(6) & 1.01\\
-2.0& 0.0390 & -0.45(8) & -0.39(6) & 0.3(4) & 1.30\\
-1.6& 0.0971 & -0.5025(5) & -0.22(2) & -0.27(3) & 0.42\\
-1.2& 0.1712 & -0.49(1) & -0.42(1) & -0.00(4) & 0.98\\
-0.8& 0.3542 & -0.496(2) & -0.418(4) & -0.05(1) & 0.32\\
-0.4& 0.5722 & -0.493(2) & -0.473(3) & 0.006(8) & 0.64\\
0.0& 1.0000 & -0.496(1) & -0.464(3) & -0.016(6) & 1.02\\
0.4& 1.4929 & -0.4974(7) & -0.450(2) & -0.029(4) & 0.70\\
0.8& 2.1795 & -0.4978(7) & -0.402(2) & -0.075(4) & 0.60\\
1.2& 2.7994 & -0.494(1) & -0.386(2) & -0.061(6) & 0.49\\
1.6& 3.6330 & -0.4986(7) & -0.331(3) & -0.138(7) & 1.09\\
2.0& 4.3998 & -0.498(1) & -0.304(4) & -0.16(1) & 2.21\\
2.4& 5.0691 & -0.4980(5) & -0.298(2) & -0.157(5) & 0.61\\
2.8& 5.8386 & -0.4975(5) & -0.278(2) & -0.166(5) & 0.53\\
3.2& 6.6344 & -0.4967(7) & -0.256(2) & -0.180(7) & 0.69\\
3.6& 7.3683 & -0.4982(6) & -0.240(2) & -0.211(6) & 0.85\\
\hline\hline
\end{tabularx}
\end{center}
\end{table}

In Fig. \ref{Fig:DensityIM} we {also show the analytic continuation of the density, in Fig. \ref{Fig:PolarizationIM} of 
the magnetization, and in Fig. \ref{Fig:ContactIM} of Tan's contact to real-valued chemical potential differences}. 
Although these functions are not periodic in $\beta h$, they remain valid 
only in the original restricted domain of $[-\pi,\pi]$. A few representative values of $\beta\mu$ are shown in Figs. 
\ref{Fig:DensityIM}, \ref{Fig:PolarizationIM}, and \ref{Fig:ContactIM}, however such analytic continuations may be performed for many 
values of $\beta\mu$ on an unrestricted domain, and the equations of state for various imaginary asymmetries may be constructed. 
Such plots for representative values of $\beta h$ at a dimensionless coupling strength of $\lambda = 1.0$ for the density, magnetization, 
and Tan's contact are shown in Section~\ref{Sec:Virial}.

One of the most interesting features we observe in our results is the behavior of the magnetization $m/n^{}_0$ 
as a function of $\beta h$ and $\beta\mu$. On the imaginary side (at least in the region studied), 
this quantity is non-monotonic in both of those variables. In particular, we note that the ordering of the curves, for different values
of $\beta \mu$, is partially inverted at large enough $\beta h$. This behavior, however, results in a perfectly ordered
set of curves on the {real-valued $(\beta h)$ side}, in a way that respects both thermodynamic stability and physical intuition.

In Fig.~\ref{Fig:ContactIM} we show two possible fits and their corresponding analytic continuations for the contact, namely 
Eqs.~(\ref{Eq:DensityFit}), (\ref{Eq:DensityFitAC}), and an alternative {function
\beq
q(\beta \hi) = \gamma\left[{1 + \afit{\eta}(\beta \hi) + \bfit{\eta}((\beta \hi)^3/\pi^2)}\right]\,, \label{eq:poltype}
\eeq
where again $\afit$ and $\bfit$ are free real-valued fit} parameters and $\gamma$ is a fixed value, as discussed previously. The analytic 
continuation $\tilde{q}(\beta h)$ is given in terms of $\tilde{\eta}(\beta h)$. While the fits on the imaginary
side {appear to be of comparable quality, they} differ enough in the details that their analytic continuation to the real side displays
noticeable discrepancies. This is particularly evident for large $\beta\mu$.
{We take this to be indicative of the limitations of our approach and it should be viewed as a warning with respect to the choice
of the fit function: The Pad\'{e} form given in Eq.~\eqref{Eq:DensityFit} effectively takes into account arbitrarily 
high powers of $\cos(\beta\hi)$, whereas the ansatz~\eqref{eq:poltype} may be viewed as a low-order approximation of the 
ansatz~\eqref{Eq:DensityFit} in powers of $\cos(\beta\hi)$ which
is expected to be valid only in the vicinity of~$\beta\hi=0$. {\it A priori}, it is difficult to judge under which conditions a low-order approximation
is justified at all. In the present case, for example, the value of the coupling does not provide a direct criterion. In fact, already the 
free Fermi gas in one dimension for $\beta{\mu}>0$ is described
by an asymptotic series in powers of $\cos(\beta\hi)$. For $\beta{\mu}\ll -1$, on the other hand, it can be shown that already a low-order approximation yields
reliable results for the free Fermi gas, see also our discussion of the virial expansion in Sect.~ \ref{Sec:Virial}.
Apparently, the strength of the coupling does not enter these arguments. A variation of the strength of the coupling is only expected to change the 
numerical values of the series coefficients associated with such an expansion in powers of $\cos(\beta\hi)$ and may therefore only effectively improve or worsen
the convergence properties of this series. Note that both limits $\beta{\mu}\gg 1$ and $\beta{\mu}\ll -1$ correspond to weak-coupling limits in
our Monte Carlo study with fixed dimensionless coupling~$\lambda$ and fixed inverse temperature~$\beta$. In the regime $|\beta{\mu}| \lesssim 1$, on the other hand,
the theory is effectively in the strongly coupled regime, see also our discussion Sect.~\ref{sec:pair}.
}

\subsection{Magnetization-to-density ratio and magnetic susceptibility}


\begin{figure}[t]
  \centering
  \includegraphics[width=\columnwidth]{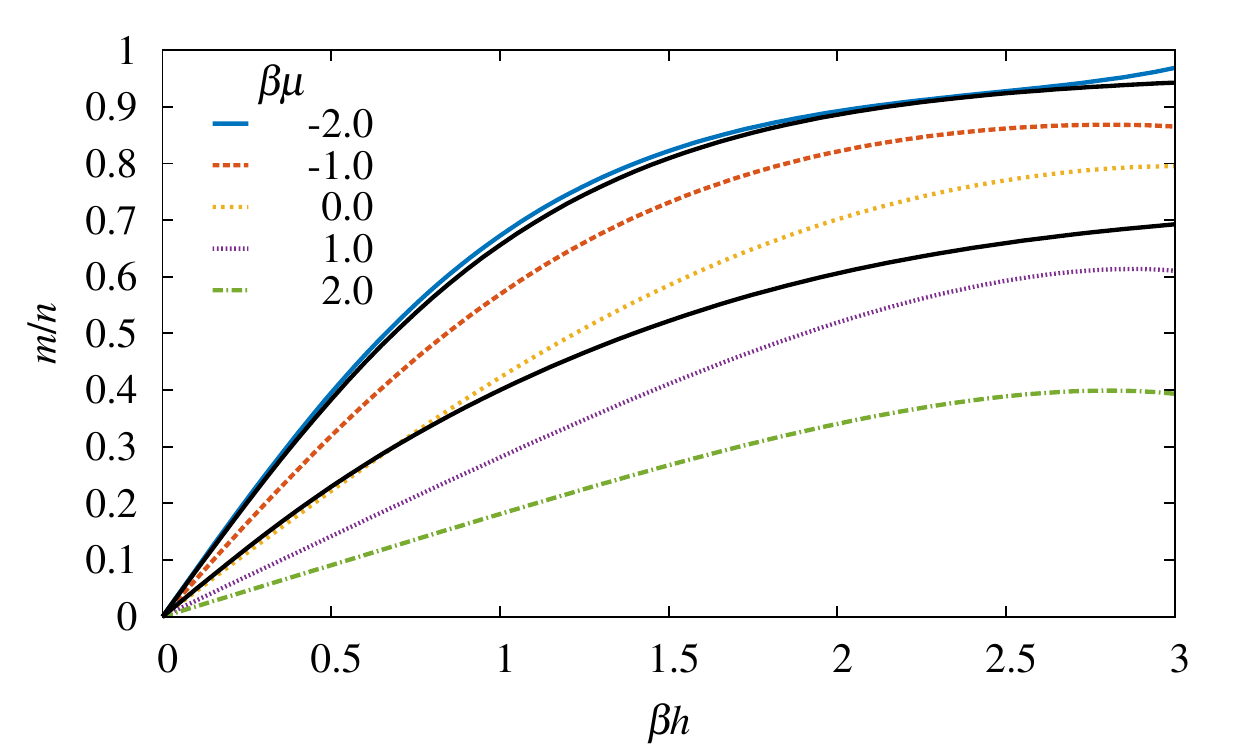}
  \caption{\label{Fig:MagnetizationDensityRatio} (Color online) Ratio of the magnetization $m$ to the density $n$ as a function of {real-valued} $\beta h$ 
  and $\beta\mu$ = -2.0, -1.0, 0, 1.0, 2.0. The solid lines show the second-order virial expansion at $\beta {\mu} = -2$ (top) and -1 (bottom). 
  Note that the virial expansion works well for $\beta \mu = -2$, but fails dramatically for $\beta \mu = -1$ and above; see Section~\ref{Sec:Virial}
  for details.}
\end{figure}


It is important to understand whether the system we are studying is appreciably magnetized in the region of parameter space that we explore here. 
To clarify this point, in Fig.~\ref{Fig:MagnetizationDensityRatio} we show the ratio of the magnetization $m$ to the density $n$. In absolute value, 
this ratio can only vary {between 0 (unpolarized) and 1 (fully polarized). 
Furthermore, it} is reassuring that $m/n$ lies within the interval $[0,1]$ after analytic continuation, 
and is a monotonically increasing function with $\beta h$. 
\begin{figure}[t]
  \centering
  \includegraphics[width=\columnwidth]{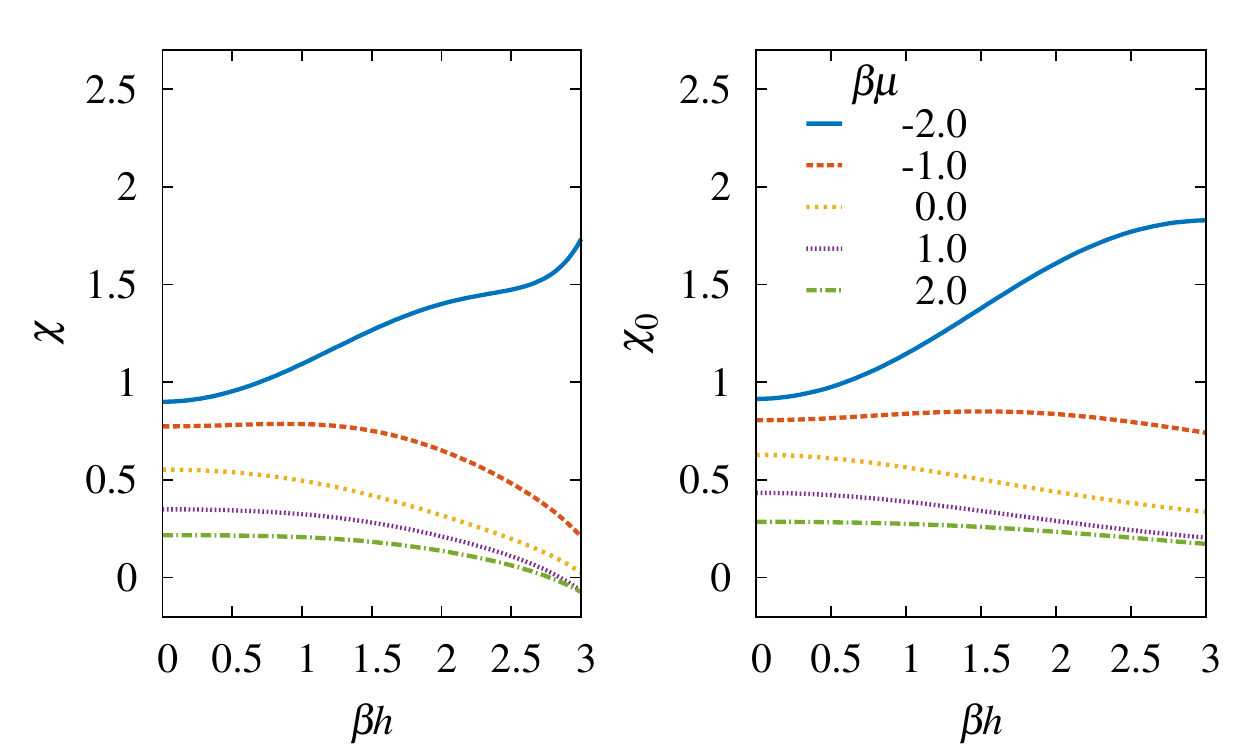}
  \caption{\label{Fig:Susceptibility} (Color online) The magnetic susceptibility $\chi$ as a function of $\beta h$ at representative values of $\beta {\mu}$, obtained by taking an analytic derivative of the polarization with respect to $\beta h$. Left panel: interacting case at $\lambda=1$. Right panel: noninteracting case in 
  the continuum.}
\end{figure}

Our results for the magnetization allow us to compute the magnetic susceptibility $\chi$ of a polarized Fermi gas simply by taking a derivative:
\beq
\chi = \frac{1}{n^{}_0}\frac{\partial m}{\partial (\beta h)}\,,
\eeq
where in practice we simply take an analytic derivative of Eq.~(\ref{Eq:PolarizationFitAC}) for each discrete value of $\beta\mu$. 
The magnetic susceptibility as a function of $\beta h$ for representative values of $\beta {\mu}$ 
at a fixed dimensionless coupling of $\lambda = 1.0$, as well as the noninteracting case is shown in Fig. \ref{Fig:Susceptibility}. Note that the deterioration in the accuracy of the analytic continuation at
large $\beta h$ (as discussed in Sec. \ref{Sec:AnalyticContinuation}) becomes more apparent as we take derivatives of physical
observables.

\section{Comparison with other approaches}\label{sec:other}

{In this section, we compare the results from our Monte Carlo simulations with those from other approaches which 
also helped us to guide the analytic continuation of our data from imaginary to real-valued chemical potential differences.
Moreover, these comparisons allow us to gain at least some insight into the phenomenology underlying one-dimensional
Fermi gases.}

\subsection{Pairing effects}\label{sec:pair}

The partition function of the noninteracting gas ($\lambda = 0$) may be computed analytically in the grand-canonical ensemble
using
\beq
\ln \mathcal{Z}(\beta\mu,\beta h) = \frac{L}{\sqrt{\pi} \lambda_T^{}} \left[ I^{}_0(z^{}_\uparrow) + I_0^{}(z^{}_\downarrow) \right]\,,\label{eq:freez}
\eeq
where
\beq
I^{}_0(z_s) =  \int_{-\infty}^{\infty} dx  \ln \left(1 + z_s {\rm e}^{-x^2} \right),\label{eq:i0}
\eeq
and $z^{}_\uparrow = {\rm e}^{\beta \mu} {\rm e}^{\beta h}$ and $z^{}_\downarrow = {\rm e}^{\beta \mu}{\rm e}^{-\beta h}$.
From this, it follows immediately that $\ln {\mathcal Z}$ can be written in terms of 
an asymptotic series of the form
\be
\ln {\mathcal Z}(\beta\mu,\beta h)  = \sum_{k=0}^{\infty}b_k{\rm e}^{k\beta\mu}\cosh(k\beta h)\,,
\ee
which, in retrospect, motivates our general forms for the ans\"atze for the fit functions used to analytically continue
the Monte Carlo data from imaginary to real-valued chemical potential differences. The coefficients $b_k$ can be related to the one-, two-, three-, $\dots$, $N$-body
problem, see also our discussion below. Note that this series converges particularly well for~$\beta\mu \ll -1$.
\begin{figure}[t]
  \centering
  \includegraphics[width=1\columnwidth]{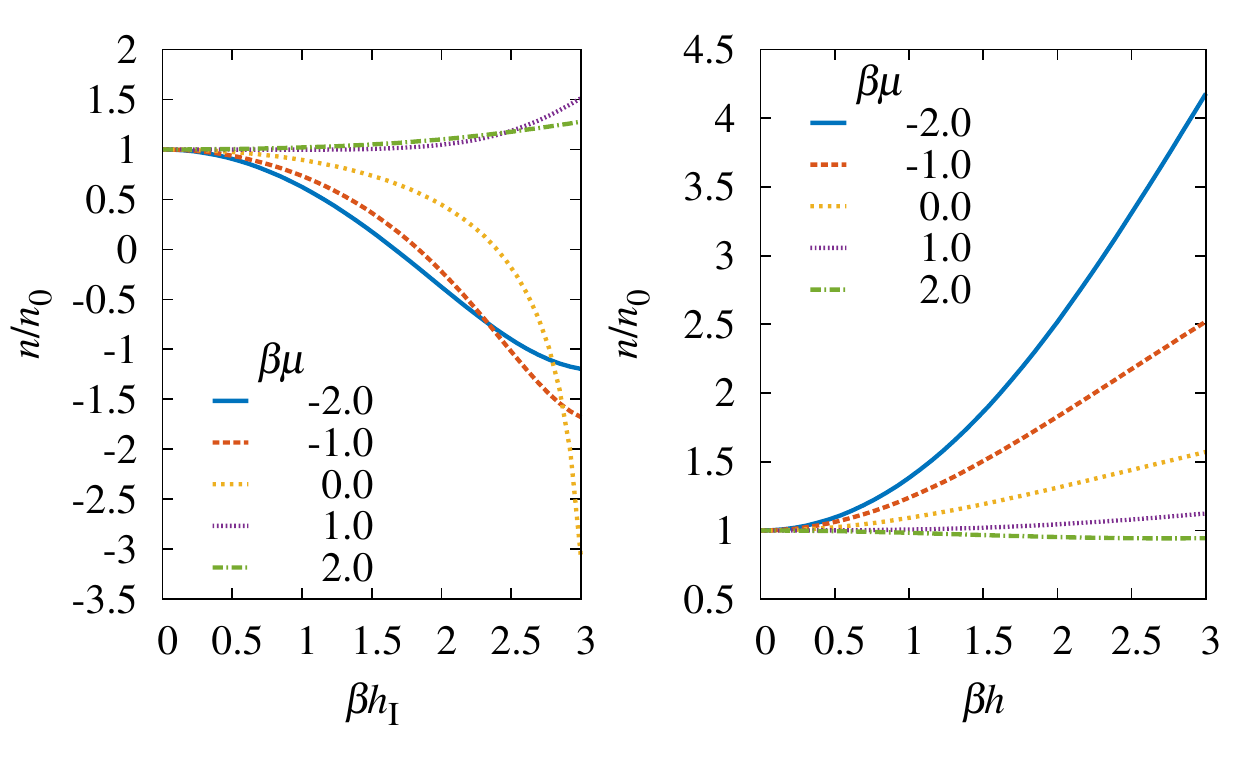}
  \includegraphics[width=1\columnwidth]{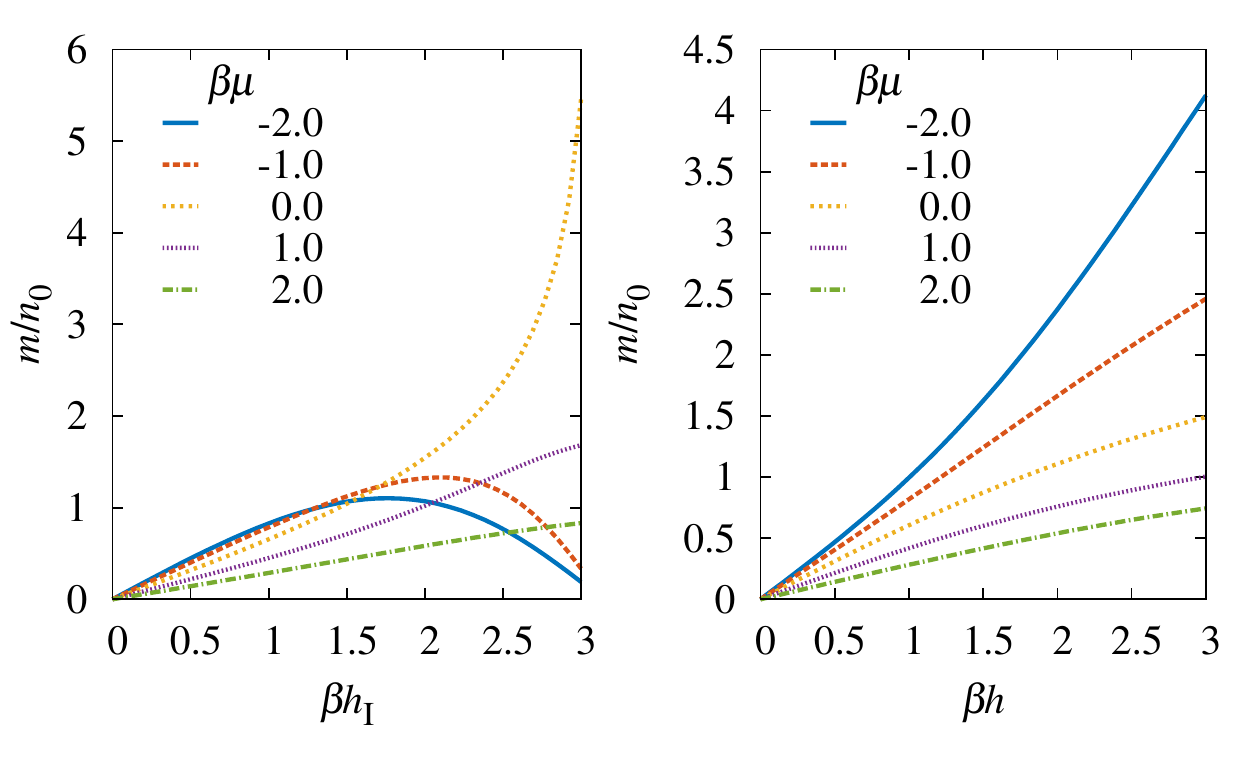}
  \caption{\label{Fig:freegas} (Color online) {Top: Density $n$ of the noninteracting Fermi gas for several values of $\beta\mu$ 
  as a function of imaginary (left panel) and real (right panel) values of $\beta h$. Bottom: Same as top panel, but showing the magnetization $m$.
  Both $n$ and $m$ are displayed in units of the density $n_0^{}$ of the noninteracting, unpolarized system.}
  }
\end{figure}

{The density and spin magnetization follows immediately from Eq.~\eqref{eq:freez}. For the density, we obtain
\beq
\label{eq_nderiv}
n \lambda^{}_T = \frac{\lambda^{}_T}{L} \frac{\partial \ln \mathcal{Z}(\beta\mu,\beta h)}{\partial (\beta \mu)} =
\frac{1}{\sqrt{\pi}} \left[I_1^{}(z^{}_\uparrow) + I_1^{}(z^{}_\downarrow) \right]\,,
\eeq
where $I^{}_1(z_s) = z_s \partial I^{}_0(z_s) / \partial z_s$, and, for the magnetization~$m$, we find
\beq
\label{eq_mderiv}
m \lambda^{}_T = \frac{\lambda^{}_T}{L} \frac{\partial \ln \mathcal{Z}(\beta\mu,\beta h)}{\partial (\beta h)} =
\frac{1}{\sqrt{\pi}} \left[I_1^{}(z^{}_\uparrow) - I_1^{}(z^{}_\downarrow) \right]\,.
\eeq
In Fig.~\ref{Fig:freegas}, we show our results for the density and the magnetization of the free Fermi gas as a function
of $\beta\hi$ for various values of~$\beta\mu$. Comparing these results for the free Fermi gas with those from our Monte Carlo
study (see Figs.~\ref{Fig:DensityIM} and~\ref{Fig:PolarizationIM}),} {we observe that both agree qualitatively, at least for finite $\beta\mu$. 
For $\beta\mu=0$, on the other hand, we observe that the results from the free gas diverge for $|\beta\hi|\to\pi$. This can
be understood from Eq.~\eqref{eq:i0}: Setting $\beta\mu=0$, we observe that the function~$I_0$ diverges
at least logarithmically in the limit $|\beta\hi|\to\pi$. We would like to add that the reason for the divergence appearing {in the results 
in} this limit can also be understood from a study of the free propagator which becomes singular 
for $\beta\mu=0$ and $\vec{p}^{\,2}=0$ in the limit $|\beta\hi|\to\pi$. Formulated in the language of thermal field theory, 
the fermionic Matsubara modes $\beta\omega_n=(2n+1)\pi$ entering the free propagator 
effectively assume the form $\beta\omega_n=2n\pi$ associated with bosonic degrees of freedom in the limit $|\beta\hi|\to\pi$, see also Ref.~\cite{BraunEtAl}. 
In other words, in this limit the fermions acquire a (thermal) zero mode. However, we emphasize that the partition function
is analytic for any finite value of $\beta\mu$.\footnote{Note that $\ln {\mathcal Z}$ is  
a sum of $I_0({\rm e}^{\beta \mu} {\rm e}^{\beta h})$ and $I_0({\rm e}^{\beta \mu} {\rm e}^{-\beta h})$ up to numerical factors.}
Therefore, no divergences occur in the limit $|\beta\hi|\to\pi$ for $\beta\mu\neq 0$.
The situation is substantially different in the case of an interacting Fermi gas as studied
with our Monte Carlo approach. Here, $\beta\mu$ is only a parameter with limited physical meaning. In fact, whereas $\mu$ determines
the energy of the free Fermi gas, the parameter~$\mu$ in the Monte Carlo study only sets the scale for the energy of the corresponding 
interacting Fermi gas. Loosely speaking,
an effective chemical potential~$\mu_{\text{eff}}$ may also be assigned to the interacting Fermi gas. In general, its value would then be different 
from the value of the parameter $\mu$. Indeed,
even for $\beta\mu=0$, the results for physical observables from our Monte Carlo study appear to be analytic as a function of $\beta\hi$. For example,
the spin magnetization $m$ diverges for $\beta\mu=0$ and $|\beta\hi|\to\pi$ in the case of the free Fermi gas, see Fig.~\ref{Fig:freegas}. 
For the interacting Fermi gas, on the other hand, $m$ appears to be analytic for $\beta\mu=0$ and only exhibits a rapid decrease in the limit~$|\beta\hi|\to\pi$, suggesting the
effective chemical potential~$\mu_{\text{eff}}$ associated with the interacting theory is small but finite, see Fig.~\ref{Fig:PolarizationIM}.}

{Although there is no spontaneous symmetry breaking in one dimension in the long-range limit, pairing of fermions is {\it a priori} still possible for any finite
coupling strength and expected to impact the ground state. It is therefore worthwhile to study pairing} in the one-dimensional Fermi gas at zero temperature. In this case, the coupling is conveniently rendered 
dimensionless with the aid of the chemical potential~$\mu$ which sets the scale:
\be
\bar{g}=\frac{\lambda}{\sqrt{|\beta\mu}|}=\frac{g}{\sqrt{|\mu|}}\,.
\ee
Thus, the scale $\beta$ drops out as it should be and the dimensionless coupling~$\bar{g}$ can now be viewed as a measure of the potential energy (measured
in terms of $g$) relative to the kinetic energy (measured in terms of $\mu$). We observe that, for fixed $\lambda$ and $\beta$
as in our case, we approach the weak-coupling limit for, e.g.,~$\beta\mu\gg 1$.
On the other hand, the theory becomes strongly coupled for fixed~$\lambda\sim{\mathcal O}(1)$ if $|\beta\mu|\lesssim 1$.

\begin{figure}[t]
  \centering
  \includegraphics[width=1\columnwidth]{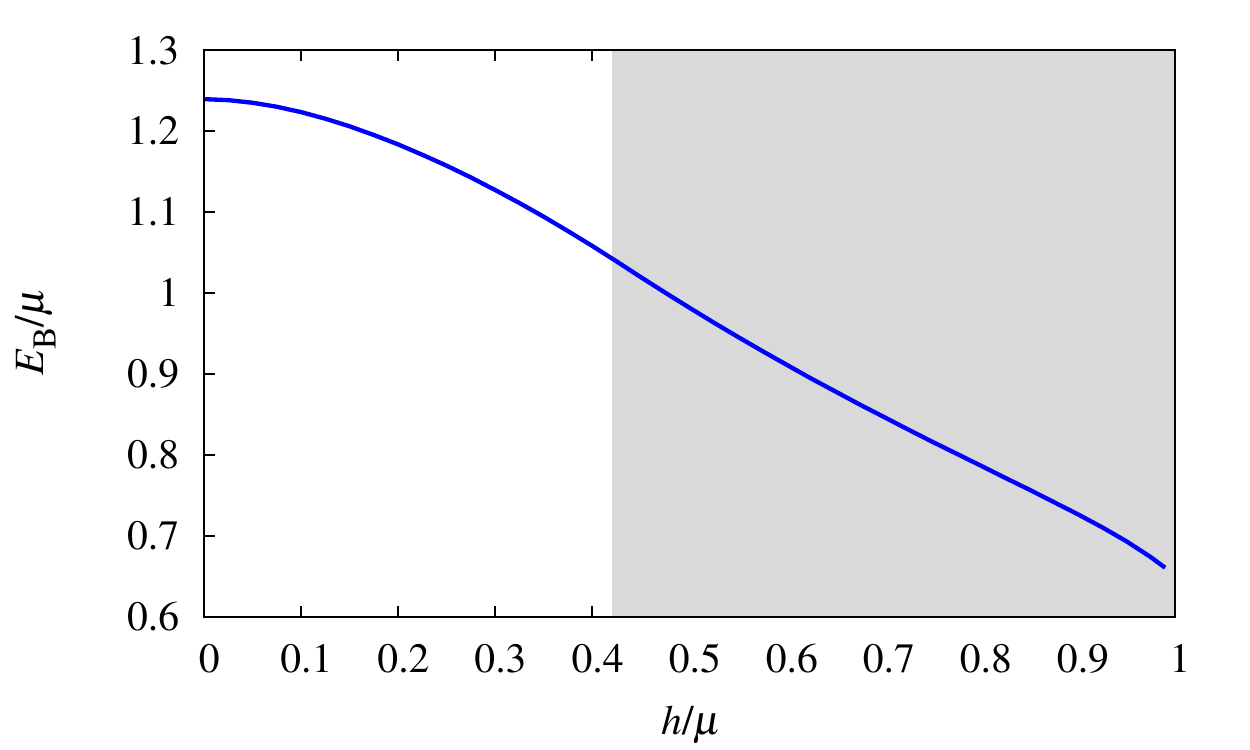}
  \caption{\label{Fig:bstate} (Color online) {Dimensionless binding energy $E_{\rm B}/\mu$ of the two-body bound state in the presence of (inert) Fermi surfaces as 
  a function of $h/\mu$ for~$\bar{g}=\pi$. The gray-shaded area depicts the regime in which the formation of a bound state with finite center-of-mass momentum is favored.}}
\end{figure}

{Keeping this in mind, let us now analyze the role of pairing effects in our Monte Carlo study by simply
considering the two-body problem in the presence of two Fermi surfaces, in close analogy to standard BCS theory~\cite{BCS}.
The underlying Schr\"odinger equation, which has proven very useful
to understand the general phase structure of imbalanced Fermi gases~\cite{Roscher:2013cma,Roscher:2015xha}, is given by
\be
\label{eq:sg}
\!\!\!\Big[ \sum_{s=\uparrow,\downarrow}\epsilon_{s}(\partial_{x_{s}})\! -\! g\delta(x_\uparrow\!-\!x_\downarrow)
+ E_{\rm B}\Big]\Psi(x_\uparrow,x_\downarrow)=0\,.
\ee
Here},~$\Psi$ is the wave-function of the bound state. The operator~$\epsilon_s$ is defined 
as~$\epsilon_s(\partial_{x})=|-(2m)^{-1}\partial_{x}^2-\epsilon_{{\rm F},s}|$ with~$\epsilon_{{\rm F},s}$ $(s=\uparrow,\downarrow)$ 
being the Fermi energy of the up- and down-fermions, respectively. Interestingly, the solution of this one-dimensional two-body problem can in principle 
also be given in closed {form~\cite{Roscher:2015xha}.} 
For our purposes, however, only the (binding) energy of the lowest-lying bound-state, which is obtained from a minimization of the energy~$E_{\rm B}$ with respect to the  
total momentum~$P$, is of particular interest. For illustration purposes, we show the energy of this state as a function of $h/\mu$ in Fig.~\ref{Fig:bstate}
{for~$\bar{g} = \pi$ (in the strong-coupling regime), corresponding to $\beta\mu = 1/\pi^2\approx0.1$ in} our Monte-Carlo study with fixed~$\lambda=1.0$. 
The gray-shaded area in Fig.~\ref{Fig:bstate} depicts the regime in which it is energetically most favorable to form a bound state with finite center-of-mass momentum. In a full
many-body treatment, the true ground state can potentially be inhomogeneous in this regime~\cite{Roscher:2013cma,Roscher:2015xha}.

{We observe that the {dimensionless binding energy $E_{\rm B}/\mu$} becomes smaller for increasing spin-imbalance~$h/\mu$. Moreover, we find that the formation of a two-body bound
state is no longer energetically favored {for $\bar{g}\lesssim 0.96$ which corresponds to $\beta\mu \gtrsim 1.08$ in} our Monte Carlo study with fixed~$\lambda=1.0$.
Loosely speaking, this suggests that, for fixed~$\lambda$, the Fermi gas undergoes a crossover from a strongly correlated to a
weakly correlated system in the limit $\beta\mu\gg 1$.}

{{A word of caution needs to be added here: Our study of bound-state formation
in the presence of (inert) Fermi surfaces is clearly only an approximation as the Fermi surfaces are smeared out at finite
temperature and coupling strength. Moreover, it has been} found that the spin-balanced $N_{\uparrow}+N_{\downarrow}$-problem
``dimerizes", i.e. the ground state energy of this system in the zero-temperature limit is given by the $(N_{\uparrow}+N_{\downarrow})/2$ times the 
binding energy of the associated two-body bound state (see, e.g., Refs.~\cite{OneDBooks,McGuire,Rammelmuller:2015zaa,2015arXiv150703174G}). 
Thus, even in the limit of small dimensionless coupling~$\bar{g}$, bound states are formed. Nevertheless, as in our study of bound-state formation
in the presence of Fermi surfaces, the dimensionless energy of the
system (i.e. energy measured in units of $\mu$) decreases for fixed coupling~$g$ and increasing~$\mu$. However, the critical coupling for bound-state 
formation turns out to be zero in the exact solution, independent of the degree of spin imbalance of the system.} 

{With respect to our Monte Carlo simulations, these considerations imply
that pairing effects are present {for all values} of $\mu$ and $h$ and are at least partly responsible for the difference between the results for the free Fermi gas
and our Monte Carlo results. It should also be added that our Monte Carlo study is bound to finite temperature $|\beta h| \leq \pi$ which hinders a direct
comparison between our {Monte Carlo} results and the exact solutions~\cite{McGuire,OneDBooks} only available for the zero-temperature limit. At finite temperature, thermal
energy is ``pumped" into the system, effectively resulting in dissociation of the bound states.\footnote{{Note that our Monte Carlo results are always
given as a function of $\beta h$. Thus, large values of $\beta h$ correspond to low temperatures for fixed chemical potential difference~$h$, whereas
small $\beta h$ is associated with high temperatures in this case.}}}

{We close by adding a comment on local ordering in one-dimensional Fermi gases. The formation of bound states can be considered
as a necessary condition for the formation of a superfluid condensate. Of course, as stated above, there is no spontaneous symmetry breaking
in these one-dimensional systems in the long-range limit. Nevertheless, the emergence of local ordering, i.e. the emergence of a condensate in the presence of 
a sufficiently large infrared cutoff, may be possible. In three-dimensional systems, these types of phases are associated with 
{precondensation~\cite{precond,Roscher:2015xha}.} In experiments, such an infrared cutoff scale is effectively set by the inverse of the length
scale associated with the confining geometry, e.g. a harmonic trap potential. Since the extent of inhomogeneous phases in the space spanned by
the experimental parameters is expected to be {large~\cite{Roscher:2013cma,inh1d}, it} may indeed be worthwhile to further study the fate of these phases at finite temperature with
the aid of our present Monte Carlo setup.}

\subsection{\label{Sec:Virial} Virial expansion}

{Let us now consider the virial expansion of the partition function, i.e. an expansion in powers of $z\equiv {\rm e}^{\beta{\mu}}$.}
In the $\beta\mu\to-\infty$ limit, where the virial expansion is valid, we can evaluate the density and the magnetization {order by order}.
Indeed, at leading order in $z$, $n^{}_{\uparrow,\downarrow}\lambda_T^{} = z^{}_{\uparrow,\downarrow}$, and therefore
\beq
 n\lambda_T^{} = (n^{}_\uparrow + n^{}_\downarrow)\lambda_T^{} = 2 {\rm e}^{\beta\mu}\cosh(\beta h)
\eeq
which leads to
\beq
\frac{n}{n^{}_0} = \cosh(\beta h)
\eeq
where $n^{}_0$ is the density for the unpolarized system. This result is the leading order in the virial expansion and 
{does not depend on the interaction}. Similarly, we find for the magnetization that
\beq
\frac{m}{n^{}_0} = \frac{n^{}_\uparrow - n^{}_\downarrow}{n^{}_0} =  \sinh(\beta h)\,,
\eeq
at leading order in $z$ {which then yields}
\beq
\frac{m}{n} = \tanh(\beta h)\,,
\eeq
which is valid at the same (leading) order in $z$.

\begin{figure}[t]
  \centering
  \includegraphics[width=\columnwidth]{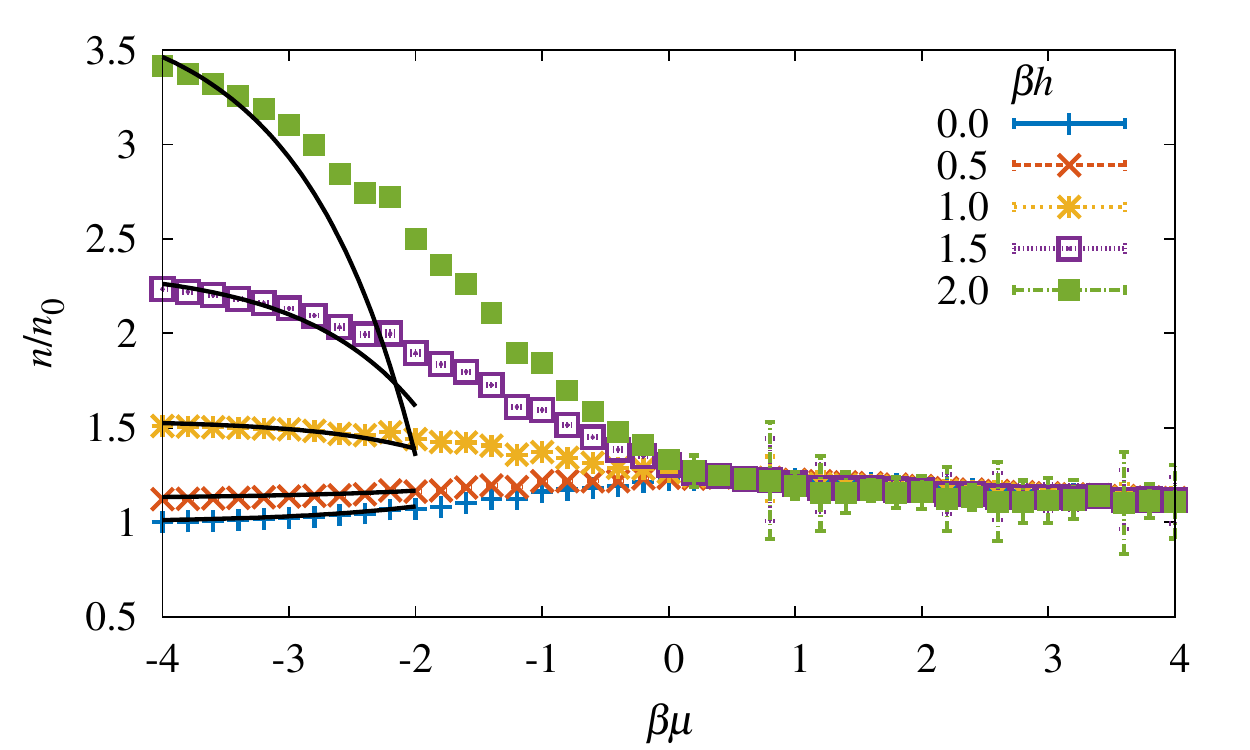}
  \includegraphics[width=\columnwidth]{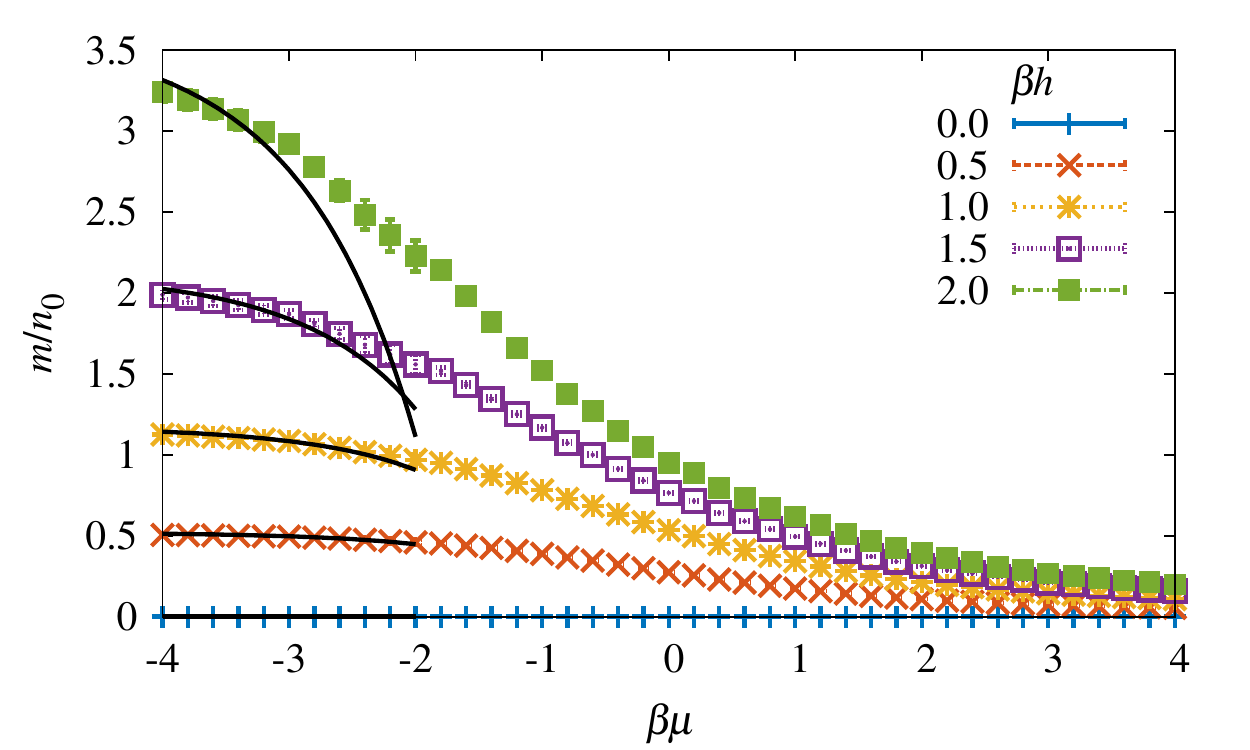}
  \caption{\label{Fig:DensityBetaMu} (Color online) Top: Analytic continuation of the density $n$ 
  (in units of its unpolarized, noninteracting counterpart $n_0^{}$) 
  as a function of $\beta\mu$ at a constant coupling $\lambda = 1.0$ and $\beta h$ = 0, 0.5, 1.0, 1.5, 2.0. 
  The solid black line shows the second-order virial expansion for each value of $\beta h$.
  Bottom: Same as top, but showing the magnetization $m$.
  Error bars were estimated by varying the fit parameters by an amount given by the uncertainty in the calculated fits.
  }
\end{figure}

\begin{figure}[t]
  \centering
  \includegraphics[width=\columnwidth]{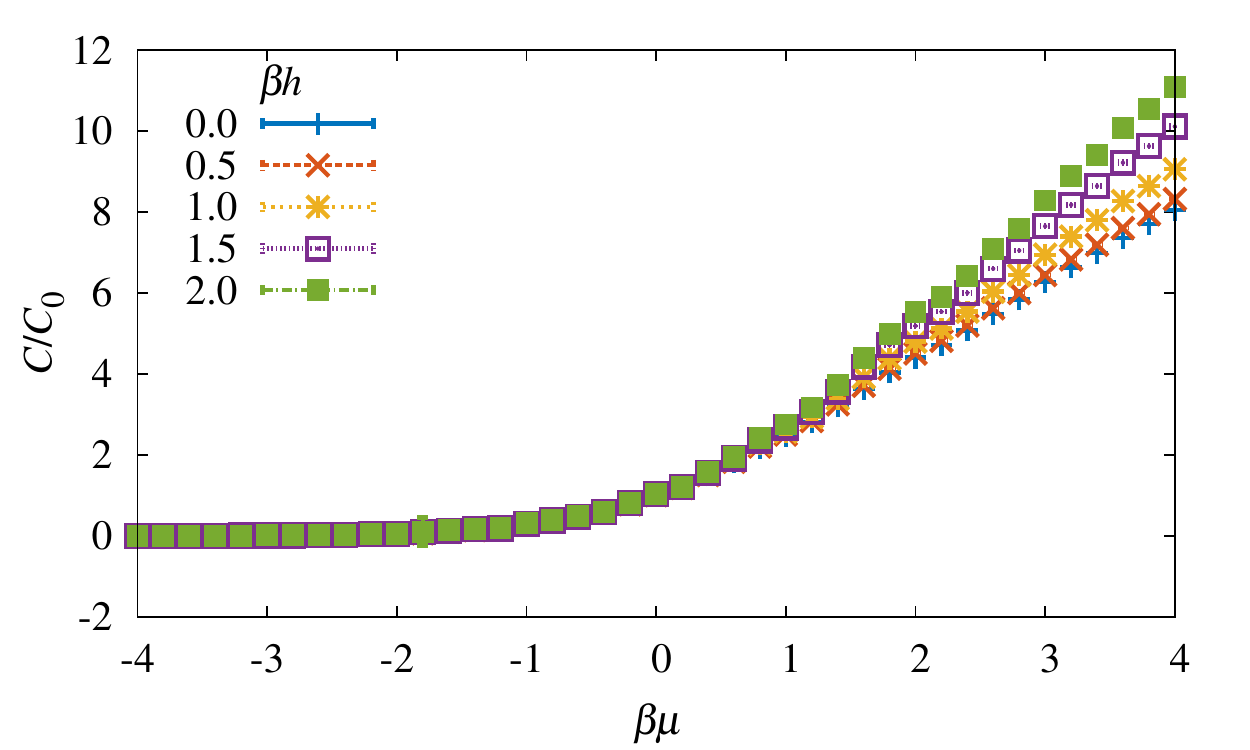}
  \caption{\label{Fig:ContactBetaMu} (Color online) Analytic continuation of Tan's contact as a function of $\beta\mu$ at a 
  constant coupling $\lambda = 1.0$ and $\beta h$ = 0, 0.5, 1.0, 1.5, 2.0. The statistical error is on the order of the size of the plotted symbol.
  $C_0^{}$ is the contact at $\beta \mu = \beta h = 0$.
  }
\end{figure}

In general, accessing higher orders in the virial expansion requires {solving the two-, three-, $\dots$, $N$-body problems 
(see, e.g., Ref.~\cite{VirialReview}).} 
The grand-canonical partition function for systems with chemical potential asymmetry may be written as
\beq
\label{eq_Zvirial2}
\mathcal{Z} = \sum_{N=0}^{\infty}\sum_{M} z^N w^M Q_{N_{\uparrow},N_{\downarrow}}\,,
\eeq
{where $N = N^{}_\uparrow+N^{}_\downarrow$ is the total particle number and $M = N^{}_\uparrow - N^{}_\downarrow$ measures the spin polarization.
Moreover, we have introduced the quatitites $w\equiv e^{\beta h}$ and $Q_{N^{}_\uparrow,N^{}_\downarrow}$ being the canonical partition  
function of a system with $N^{}_\uparrow$ spin-up fermions and $N^{}_\downarrow$ spin-down fermions. 
Expanding Eq.~(\ref{eq_Zvirial2}) to {second order} in~$z$ yields
\be
\!\!\!\!\!\!\mathcal{Z} &=& 1+2Q^{}_{1,0}z\cosh(\beta h) \nonumber\\
&& +\, Q^{}_{1,1}z^2+2Q^{}_{2,0}z^2\cosh(2\beta h) + {\mathcal O}(z^3)\,,
\ee
where $Q_{1,0} = L/\lambda_T$ and $Q_{1,1}$ and $Q_{2,0}$ may be determined by direct calculation~\cite{EOS1D}. Note that the expansion
coefficients $Q_{N_{\uparrow},N_{\downarrow}}$ depend implicitly on the coupling~$\lambda$.} The density $n$ and magnetization $m$ in the 
second-order virial expansion may then be determined from $\mathcal{Z}$ through 
Eqs.~(\ref{eq_nderiv}) and (\ref{eq_mderiv}) and compared with Monte Carlo results in the $z\rightarrow0$ limit. 
We do this explicitly in Fig.~\ref{Fig:DensityBetaMu} and find excellent agreement for small $z$ and $\beta h$; however, the quality 
of the agreement deteriorates quickly as $\beta h$ is increased {unless $z \simeq 0$, as expected.}
For completeness, Fig.~\ref{Fig:ContactBetaMu} shows Tan's contact as a function of $\beta {\mu}$, for several values of $\beta h$,
and at $\lambda=1$.

{ 
\subsection{\label{Sec:Perturbation} Lattice perturbation theory}

As an additional verification of the equations of state obtained through an analytic continuation, we performed a next-to-leading order
lattice perturbation theory calculation by expanding the lattice grand-canonical partition function $\mathcal{Z}(\beta\mu,\beta h)$ in the 
dimensionless parameter $A \equiv \sqrt{2(e^{\tau g}-1)}$ (which arises naturally in lattice calculations, see Ref.~\cite{MCReviews})
about the noninteracting limit ($A=0$), up to the second non-vanishing term. Such an expansion on the lattice yields
\beq
\label{Eq:PartitionFunctionPT}
\ln\mathcal{Z} = \ln\mathcal{Z}_0 + \frac{\beta}{2 \tau N_x} (Ae^{\beta\mu})^2 \prod_{s=\uparrow,\downarrow} \left(\sum_{k} \frac{e^{-\beta\epsilon^{}_k}}{
1+z_s e^{-\beta\epsilon^{}_k}}\right), 
\eeq
where $\mathcal{Z}_0(\beta\mu,\beta h)$ is the partition function of the noninteracting gas, $\epsilon^{}_k = k^2/2m$ and the
sum over $k$ is over all possible lattice momenta. The density $n/n_0$ and magnetization $m/n_0$ in terms of this perturbation
theory therefore follow from Eq. (\ref{Eq:PartitionFunctionPT}) using Eqs. (\ref{eq_nderiv}) and (\ref{eq_mderiv}). We display the results of
this analytic calculation with the numerical Monte Carlo results and the equations of state of the free gas in 
Fig.~\ref{Fig:PTComparison}.
\begin{figure}[t]
  \centering
  \includegraphics[width=\columnwidth]{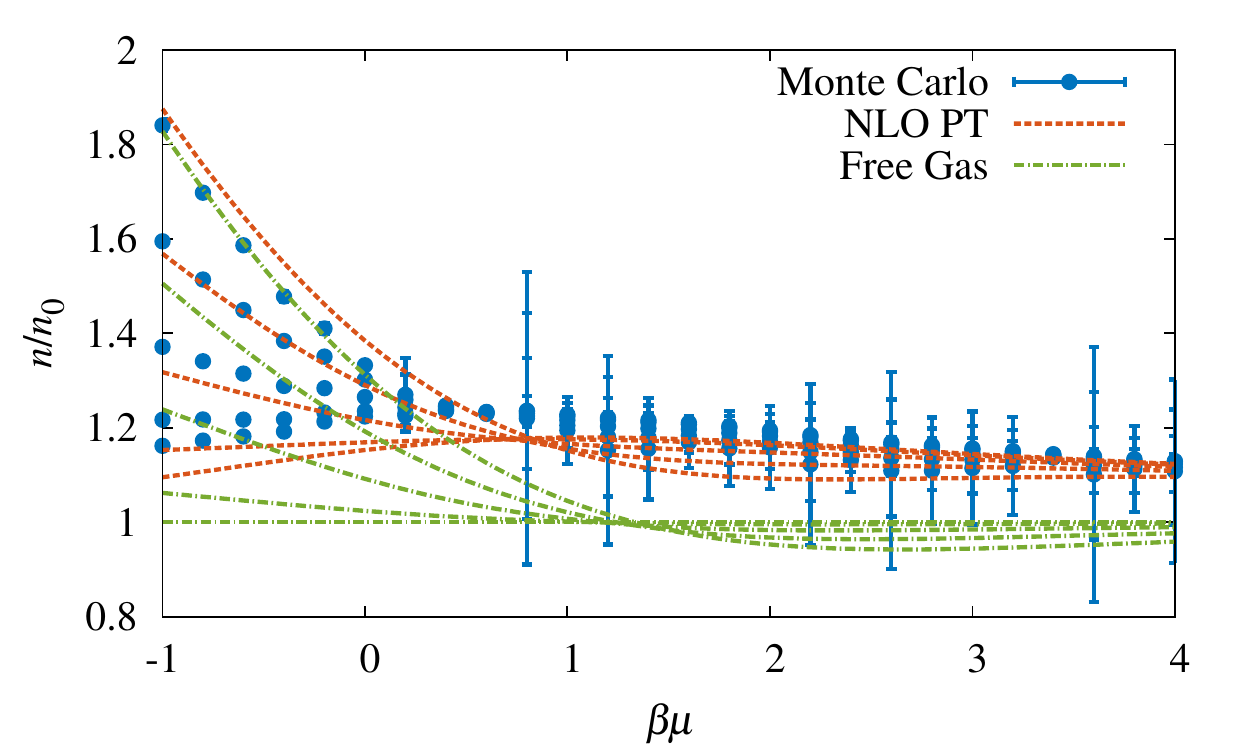}
  \includegraphics[width=\columnwidth]{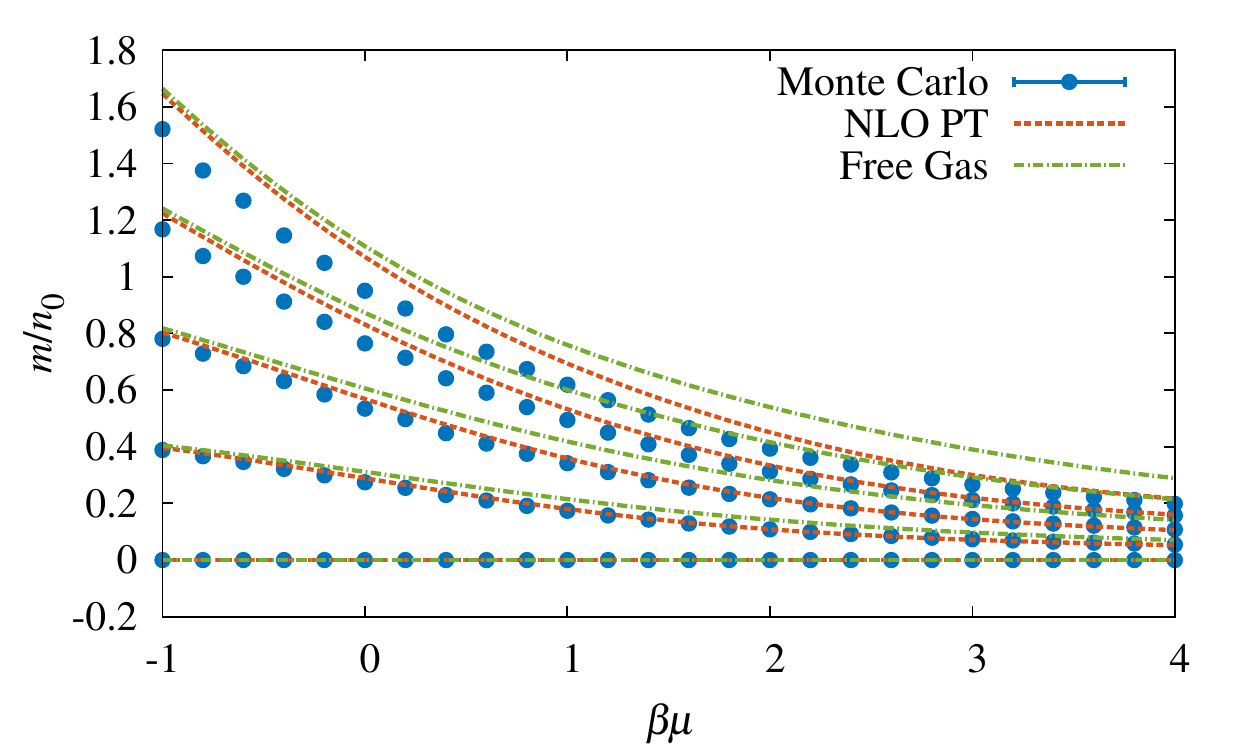}
  \caption{\label{Fig:PTComparison} (Color online) Comparison of the Monte Carlo results after the analytic continuation, a next-to-leading order perturbation theory calculation (NLO PT), and the free gas for the density (top) and the magnetization (bottom). Results are displayed for $\beta h = 0.0, 0.5, 1.0, 1.5$ and 2.0  (bottom to top) in the strongly interacting region $\beta\mu > -1$.
  }
\end{figure}
In all cases the results obtained with our proposed analytic-continuation approach lie generally between the leading (noninteracting) calculation and
the next-to-leading-order result.
}
\section{Summary and conclusions} 

In summary, we have performed a non-perturbative characterization of the density $n$, magnetization $m$, magnetic
susceptibility $\chi$, and Tan's contact $C$ of a 1D, attractively interacting Fermi gas. To this end, we implemented the conventional 
finite-temperature lattice Monte Carlo formalism, but generalized to include complex chemical potentials. When the chemical potential
asymmetry $h$ is purely imaginary, there is no sign problem and the Monte Carlo calculation can be carried out as usual.
Our Monte Carlo results on the imaginary-$h$ side are therefore exact up to controlled (statistical and systematic) uncertainties.  

To obtain results for real $h$, we performed fits to our numerical data and implemented an analytic continuation, i.e. we set $h \to ih$.
In some regions of parameter space, different functional forms for the fits may yield very different analytic continuations. However,
very simple functional forms such as polynomials can be discarded as they are likely too simple to capture {the essential physics}. Generally
speaking, at low enough $\beta h$ all fits lead to (approximately) the same analytic continuation. As we show in 
Fig.~\ref{Fig:MagnetizationDensityRatio}, however, even for low $\beta h$, we achieve non-trivial magnetization ratios
as large as $m/n \simeq 0.3-0.5$.

We have presented our results as a function of the dimensionless parameters $\beta\mu$ and $\beta h$, but focused on
an intermediate- to strong-coupling {regime $\lambda = \sqrt{\beta} g \sim {\mathcal O}(1)$ as a non-trivial case of relevance for future studies.
Our results for $n$ and $m$ agree qualitatively with the 
free Fermi gas where differences may be traced back to pairing effects. Moreover, our results
are consistent with the second-order virial expansion, which is non-perturbative in the interaction,
in the regime $\beta\mu < 0$, where that expansion is valid.}
We also note, however, that the virial expansion deteriorates as $\beta h$ is increased (at fixed $z$).

The calculations carried out in this work correspond to fixed lattice volume $L = N_x^{} \ell = 61$ and extent of the imaginary-time 
direction $\beta = N_\tau^{} \tau = 8.0$, where $\ell = 1$ and $\tau = 0.05$.
The associated systematic effects should, in principle, be further investigated, although the results of our previous work~\cite{EOS1D}
indicate that those effects are below 10\%. {We consider our present work as a proof-of-principle study of our imaginary spin-imbalance approach. However,
results for, e.g., the thermal equations of state, could already be extracted from it and compared to present and future experiments~\cite{Experiments1D}, if available.
In any case, our present study is mostly aimed at paving the way to more computationally demanding systems in two and three dimensions. For example, our approach
allows one to map out to some extent the finite-temperature phase diagram 
of spin-imbalanced unitary Fermi gases in three dimensions, and therefore permits one to, 
at least, narrow down the regime in parameter space in which the critical point
is located.} In that regard, all of our present results indicate that calculations in higher dimensions should be feasible with the 
proposed method.

\acknowledgments
We thank W. J. Porter, and L. Rammelm\"uller for useful discussions.
J.B. and D.R. acknowledge support by the DFG under Grant BR 4005/2-1 and
by HIC for FAIR within the LOEWE program of the State of Hesse. 
This material is based upon work supported by the 
National Science Foundation Graduate Research Fellowship Program under Grant No. DGE{1144081}, 
National Science Foundation Nuclear Theory Program under Grant No. PHY{1306520},
National Science Foundation Computational Physics Program under Grant No. PHY{1452635}, 
and 
National Science Foundation REU Sites Program under Grant No. ACI{1156614}.

\appendix
\begin{figure}[t]
  \centering
  \includegraphics[width=\columnwidth]{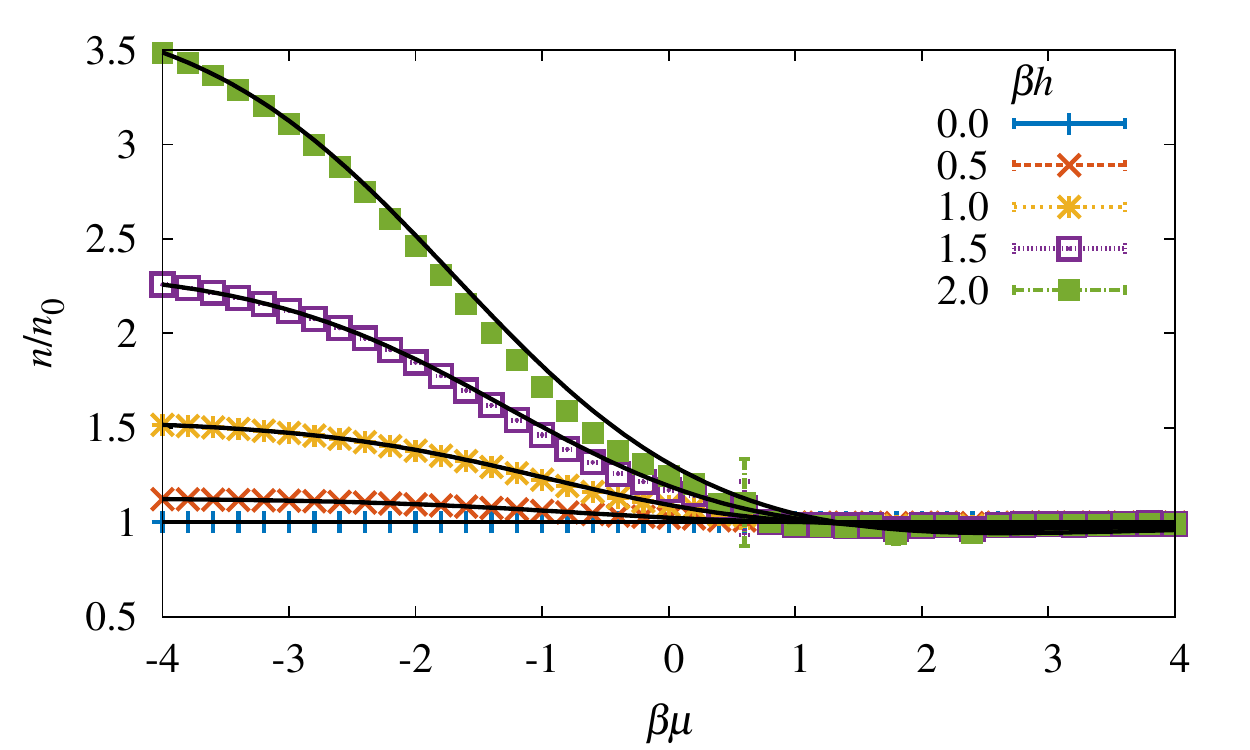}
  \includegraphics[width=\columnwidth]{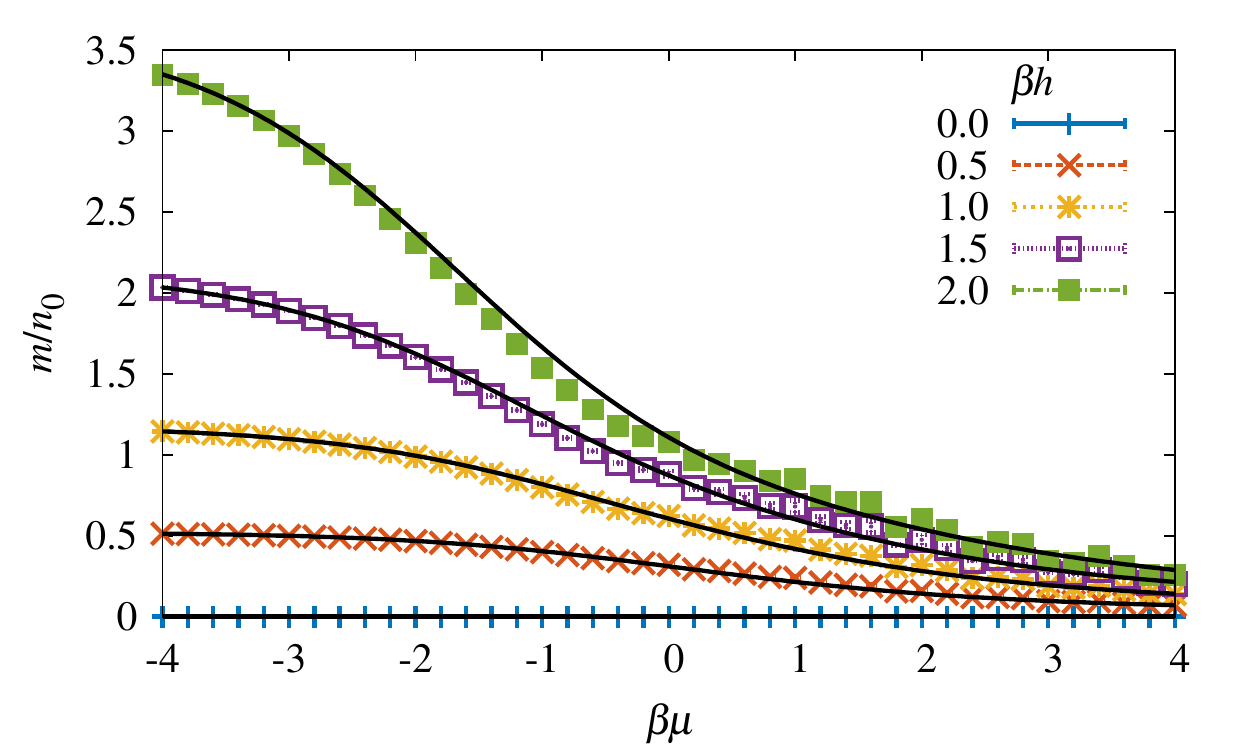}
  \caption{\label{Fig:NoninteractingComparison} (Color online) Analytic continutation of the density (top) and the magnetization (bottom) for the non-interacting system as a function of $\beta\mu$ for various values of $\beta h$. The exact solutions in the continuum limit are shown as solid black lines.}
\end{figure}

\begin{table}[]
\begin{center}
\caption{\label{Table:AltCouplingsDensityParameters}
Fit parameters for the density at alternative coupling strengths for
various values of $\beta \mu$, as well as the $\chi^2$ per degree of freedom
for each fit. Note that $\gamma$ is not a fit parameter (see main text).
The value in parentheses indicates the calculated uncertainty of the least significant digit for each fit parameter.}
\begin{tabularx}{\columnwidth}{@{\extracolsep{\fill}}c c c c c c c}
\hline\hline
$\lambda$ & $\beta\mu$ & $\gamma$ & ${\afit}$ & ${\bfit}$ & ${\cfit}$ & $\chi^2$\\
\hline
0.5 & -2.0 & 1.0272 & -0.8876(3) & -0.1553(7) & -0.0029(7) & 1.45\\
0.5 &-1.0 & 1.0624 & -0.7412(8) & -0.299(2) & -0.027(2) & 2.63\\
0.5 & 0.0 & 1.0937 & -0.5216(4) & -0.392(3) & -0.064(4) & 2.21\\
0.5 & 1.0 & 1.1018 & -0.2(3) & -0.2(3) & -0.05(6) & 0.73\\
0.5 & 2.0 & 1.0908 & -0.20(2) & -0.23(2) & -0.014(2) & 0.64\\
2.0 & -2.0& 1.2530 & -0.712(2) & -0.163(7) & -0.007(9) & 0.69\\
2.0 & -1.0& 1.5051 & -0.483(6) & -0.278(8) & 0.11(6) & 1.97\\
2.0 & 0.0& 1.6588 & -0.502(2) & -0.491(2) & -0.012(1) & 0.98\\
2.0 & 1.0& 1.5738 & -0.35(5) & -0.37(5) & 0.006(1) & 1.78\\
2.0 & 2.0& 1.4263 & -0.17(5) & -0.18(5) & 0.005(1) & 1.14\\
\hline\hline
\end{tabularx}
\end{center}
\end{table}

\begin{table}[t]
\begin{center}
\caption{\label{Table:AltCouplingsMagnetizationParameters}
Fit parameters for the magnetization at alternative coupling strengths for
various values of $\beta \mu$, as well as the $\chi^2$ per degree of freedom
for each fit.
The value in parentheses indicates the calculated uncertainty of the least significant digit for each fit parameter.}
\begin{tabularx}{\columnwidth}{@{\extracolsep{\fill}}c c c c c c}
\hline\hline
$\lambda$ & $\beta\mu$ & ${\afit}$ & ${\bfit}$ & ${\cfit}$ & $\chi^2$\\
\hline
0.5 & -2.0 & 1.082(2) & 0.193(2) & 0.000(3) & 0.72\\
0.5 & -1.0 & 1.170(3) & 0.464(3) & 0.020(3) & 3.84\\
0.5 & 0.0 & 1.109(3) & 0.778(2) & 0.094(2) & 26.24\\
0.5 & 1.0 & 0.770(9) & 0.708(9) & 0.186(8) & 756.92\\
0.5 & 2.0 & 0.487(7) & 0.64(1) & 0.25(1) & 458.75\\
2.0 & -2.0 & 1.069(5) & 0.179(4) & 0.010(6) & 0.51\\
2.0 & -1.0 & 1.023(8) & 0.392(7) & 0.027(8) & 1.57\\
2.0 & 0.0 & 0.719(4) & 0.539(4) & 0.049(5) & 6.39\\
2.0 & 1.0 & 0.422(2) & 0.562(3) & 0.093(4) & 9.16\\
2.0 & 2.0 & 0.265(1) & 0.586(3) & 0.115(3) & 16.25\\
\hline\hline
\end{tabularx}
\end{center}
\end{table}

\begin{figure}[b]
  \centering
  \includegraphics[width=\columnwidth]{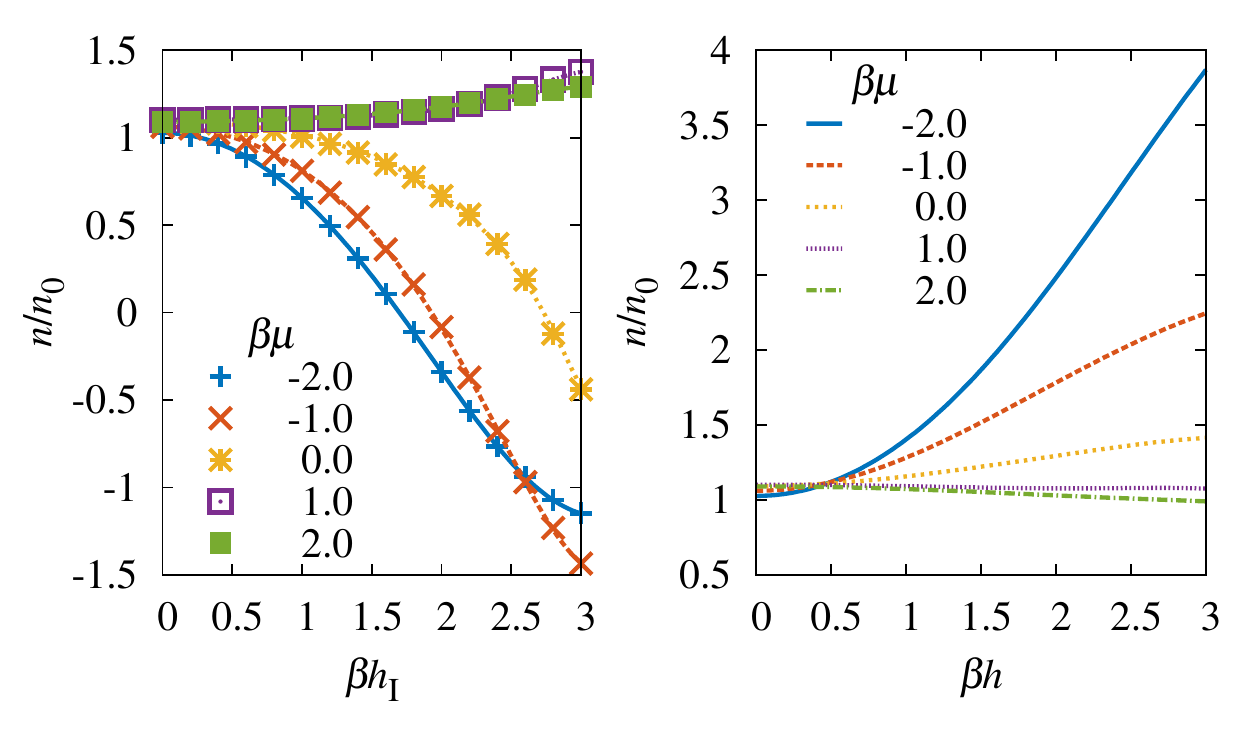}
  \includegraphics[width=\columnwidth]{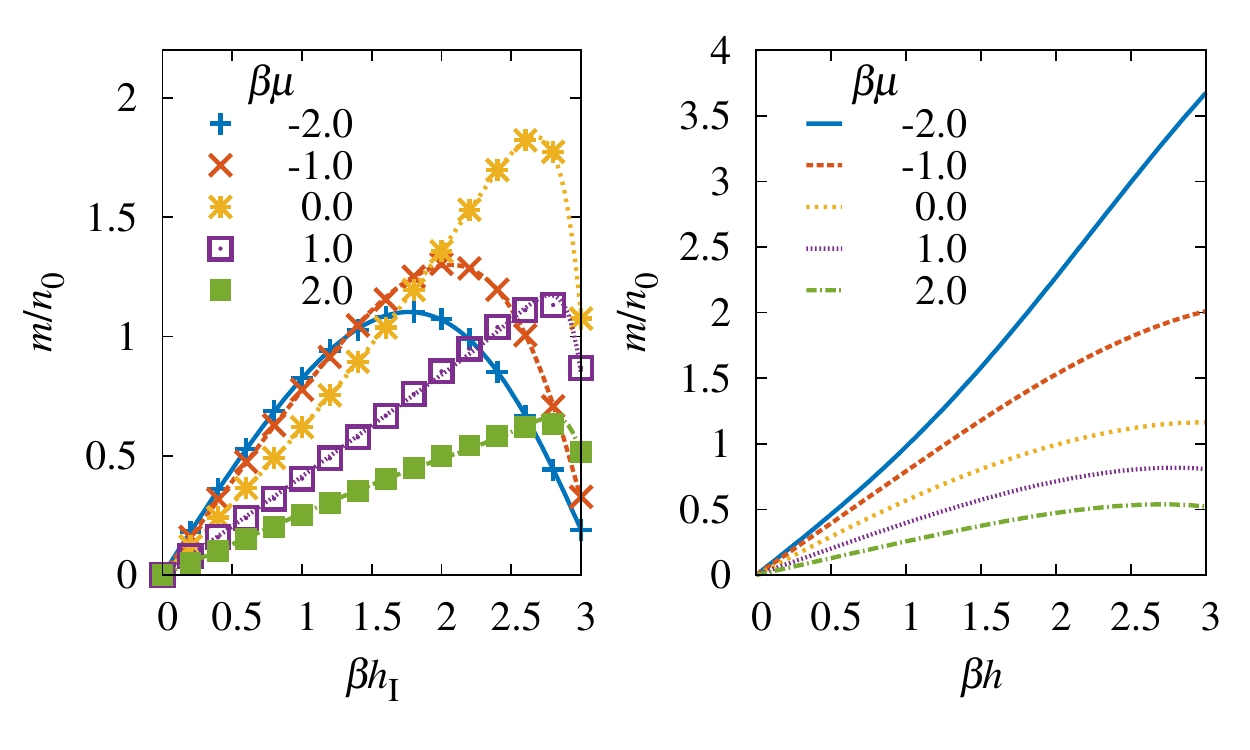}
  \caption{\label{Fig:Lambda05EOS} (Color online) Top: Density at a decreased coupling strength of $\lambda = 0.5$ as a function of the imaginary chemical potential difference $\beta\hi$ (left) and its analytic continuation to $\beta h$ (right). Bottom: Magnetization at a decreased coupling strength of $\lambda = 0.5$ as a function of the imaginary chemical potential difference $\beta\hi$ (left) and its analytic continuation to $\beta h$ (right).}
\end{figure}

\begin{figure}[]
  \centering
  \includegraphics[width=\columnwidth]{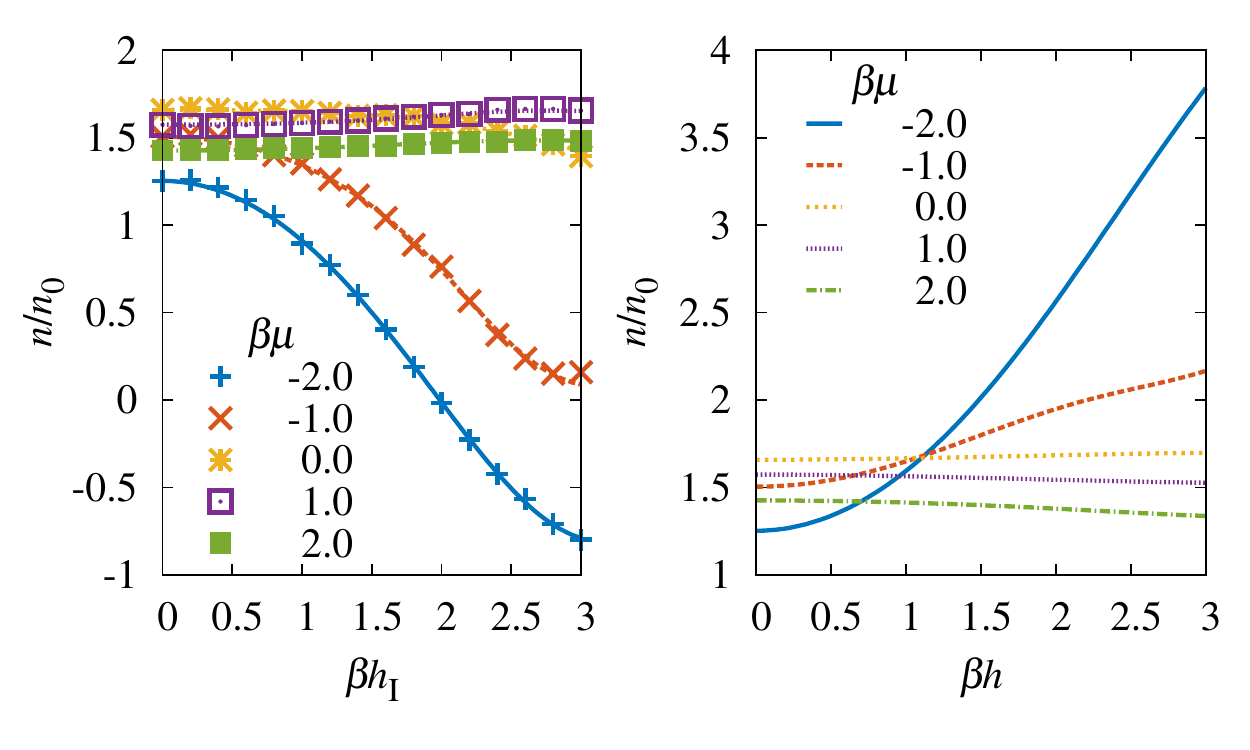}
  \includegraphics[width=\columnwidth]{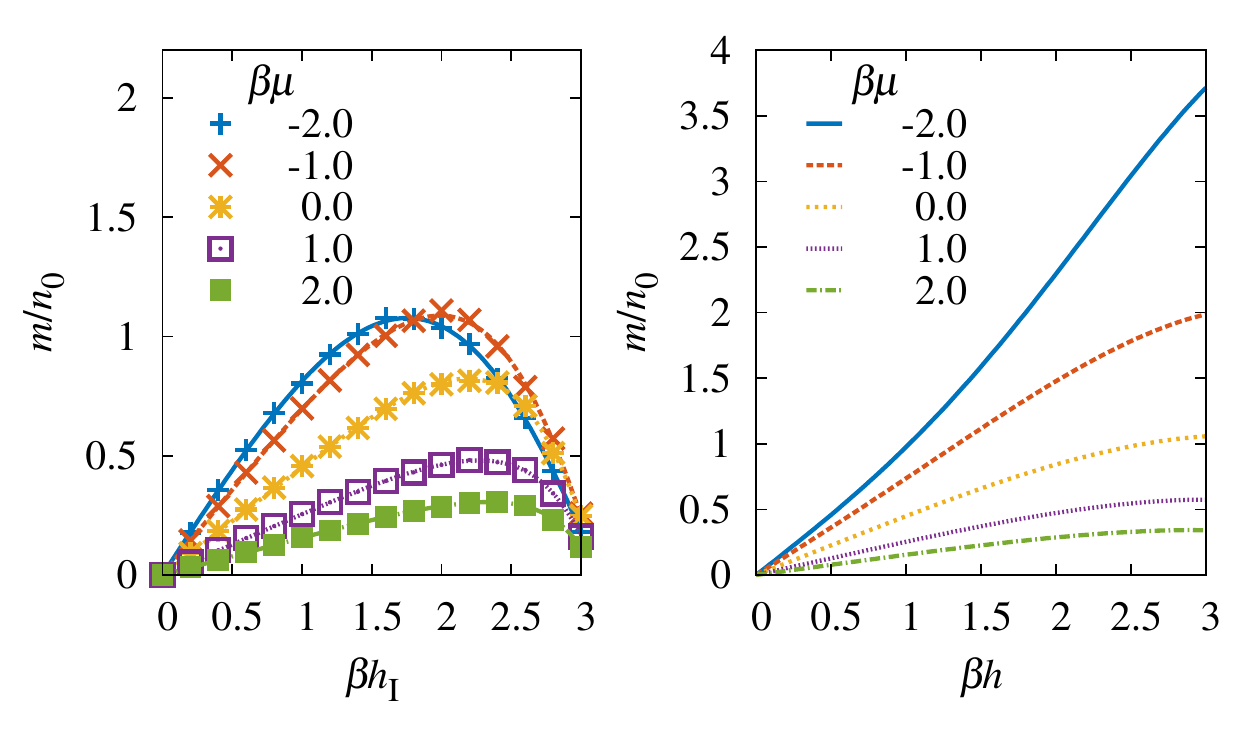}
  \caption{\label{Fig:Lambda2EOS} (Color online) Top: Density at an increased coupling strength of $\lambda = 2.0$ as a function of the imaginary chemical potential difference $\beta\hi$ (left) and its analytic continuation to $\beta h$ (right). Bottom: Magnetization at a increased coupling strength of $\lambda = 2.0$ as a function of the imaginary chemical potential difference $\beta\hi$ (left) and its analytic continuation to $\beta h$ (right).}
\end{figure}

{ 
\section{Analytic continuation of the non-interacting system}
In order to verify that the approach of analytically continuing the equations of state of the density $n/n_0$ and magnetization
$m/n_0$ is valid using {the fit ans\"atze developed} in the main text, we have performed the same prescription for the non-interacting polarized
Fermi gas and compared with the exact solution for these quantities in the continuum limit. The results comparing the Monte Carlo
and analytic solutions are shown in Fig. \ref{Fig:NoninteractingComparison}. There is excellent agreement between the {two calculations
within systematic and} statistical error, which demonstrates a measure of validity for the analytic continution in the interacting case.
In the limit of $\lambda \rightarrow 0$ on the imaginary side, $m(\beta\hi)/n_0$ converges to a sawtooth wave of periodicity $2\pi$ for 
large positive $\beta\mu$, whose behavior in the vicinity of $\beta\hi = \pm \pi$ is difficult to capture for the given ansatz. As such, the
determined fit parameters $\afit$, $\bfit$, and $\cfit$ are not necessarily smooth functions of $\beta\hi$, and noise is introduced into the result 
for $m(\beta\mu)/n_0$. The smoothness of the interacting results shown in the main text for $\lambda > 0$ significantly reduces such effects.

\section{Equations of state for alternative coupling strengths}
In the main text we have illustrated the method of calculating the equation of state of polarized,
interacting fermions using complex chemical potentials at a constant coupling of $\lambda = 1.0$
for clarity and as a proof of principle. This same method can be applied to other values of the
coupling as well, considering that larger values of $\lambda$ may require a larger number of
Monte Carlo samples in order to mantain the statistical quality of the data. In Figs. \ref{Fig:Lambda05EOS}
and \ref{Fig:Lambda2EOS} we show the density and magnetization as functions of both $\beta h$ and
$\beta\hi$ for dimensionless couplings $\lambda = 0.5$ and 2.0. The fit parameters for a chosen
set of $\beta\mu$ at these couplings are provided in Tables \ref{Table:AltCouplingsDensityParameters}
and \ref{Table:AltCouplingsMagnetizationParameters}.}

\bibliographystyle{h-physrev3}

\end{document}